\documentclass[11pt]{article}
\usepackage{amsmath}
\usepackage{amssymb}
\usepackage{mathrsfs}
\usepackage{cite}
\newtheorem{Theorem}{Theorem}[section]
\newtheorem{Proposition}[Theorem]{Proposition}

\title{\textbf{Lax pairs, recursion operators and bi-Hamiltonian representations of (3+1)-dimensional Hirota type equations}}
\author{M. B. Sheftel $^1$ and D. Yaz{\i}c{\i}$^2$\\
$^1$ Department of Physics, Bo\u{g}azi\c{c}i University, Bebek\\ 34342 Istanbul,  Turkey \\
$^2$ Department of Physics, Y{\i}ld{\i}z Technical University,\\ Esenler, 34220 Istanbul, Turkey\\
\vspace*{2mm}
E-mail: mikhail.sheftel@boun.edu.tr, yazici@yildiz.edu.tr}
\date{}
\begin{document}
\maketitle

\begin{abstract}
 We consider (3+1)-dimensional second-order evolutionary PDEs where the unknown $u$ enters only in the form of the 2nd-order partial derivatives. For such equations which possess a Lagrangian, we show that all of them have a symplectic Monge--Amp\`ere form and determine their Lagrangians. We develop a calculus for transforming the symmetry condition to a ``skew-factorized'' form from which we immediately extract Lax pairs and recursion relations for symmetries, thus showing that all such equations are integrable in the traditional sense. We convert these equations together with their Lagrangians to a two-component form and obtain recursion operators in a $2\times 2$ matrix form. We transform our equations from Lagrangian to Hamiltonian form by using the Dirac's theory of constraints. Composing recursion operators with the Hamiltonian operators we obtain the second Hamiltonian form of our systems, thus showing that they are bi-Hamiltonian systems integrable in the sense of Magri. By this approach, we obtain five new bi-Hamiltonian multi-parameter systems in (3+1) dimensions.
\end{abstract}

\section{Introduction}

We study (3+1)-dimensional equations of the evolutionary Hirota type
\begin{equation}
  F = f - u_{tt}g = 0\quad \Longleftrightarrow\quad u_{tt} = \frac{f}{g},\qquad g\ne 0
 \label{F}
\end{equation}
where $f$ and $g$ are functions of second derivatives $u_{t1},u_{t2},u_{t3},u_{11},u_{12},u_{13}$, $u_{22},u_{23},u_{33}$ of the unknown $u$.
Here $u=u(t,z_1,z_2,z_3)$ and the subscripts denote partial derivatives of $u$, such as $u_{ij} = \partial^2 u/\partial z_i\partial z_j$, $u_{ti} = \partial^2 u/\partial t \partial z_i$. Equations of this type arise in a wide range of applications including non-linear physics, general relativity, differential geometry and integrable systems. The examples are heavenly equations of Pleba\'nski arising in the theory of self-dual gravity \cite{pleb}.
These examples present a motivation for the evolutionary Hirota type equation to be taken in the form \eqref{F} instead of just $u_{tt} = f$ with $g=1$.
The factor $g$ plays the role of integrating factor of the variational calculus, i.e. when equation \eqref{F} is of the Euler--Lagrange form, the equation
$u_{tt} = f/g$ may not be of this form but it will still possess a Lagrangian.

We study here Lax pairs, recursion operators and bi-Hamiltonian structures of the integrable equations of the form \eqref{F}.

We start with the description of all equations \eqref{F} which possess a Lagrangian. We show that Lagrangian evolutionary equations of the Hirota type have a symplectic Monge-Amp\`ere form and we find their Lagrangians. Then we convert these equations to a two-component form. We apply the method which we used earlier \cite{nns,nsky,sy,sym,sy_tri,3D} for constructing a degenerate Lagrangian for two-component evolutionary form of the equation and using Dirac's theory of constraints \cite{dirac} in order to obtain Hamiltonian form of the system.

 We present the symmetry condition for one-component form in terms of  certain three-index first-order linear differential operators $L_{ij(k)}$. Our starting point is to convert the symmetry condition into a ``skew-factorized'' form from which we immediately extract Lax pair and recursion relations for symmetries
  in terms of these operators. This approach extends the method of A. Sergyeyev \cite{Artur} for constructing recursion operators where one constructs the recursion operator from a special  Lax pair built, in its turn, from the original Lax pair which should be previously known for the equation under study. On the contrary, we construct such a special Lax pair and recursion relations using the skew-factorized form of the symmetry condition rather than a previously known Lax pair. This approach is illustrated by well-known examples of heavenly equations listed, e.g. in \cite{ferdub}. We develop a general approach for deriving skew-factorized form of the symmetry condition based on some properties of the operators $L_{ij(k)}$. We obtain five different multi-parameter equations possessing skew-factorized symmetry condition together with the operators $A_i, B_i$ which are the building blocks for Lax pairs and recursions. To illustrate the general procedure, we consider in detail recursions for the first of the obtained equations, convert it into  a two-component form and end up with a recursion operator in a $2\times 2$ matrix form for the corresponding evolutionary two-component system. Composing the recursion operator with first Hamiltonian structure we end up with the second Hamiltonian structure, thus showing that our system is a new bi-Hamiltonian system. Then, skipping the details of the proofs, we repeat this procedure for other four equations possessing the symmetry condition in the skew-factorized form, thus discovering further four new bi-Hamiltonian multi-parameter systems.

The paper is organized as follows.
In section \ref{Lagrange}, we show that all Lagrangian equations \eqref{F} have the symplectic Monge--Amp\`ere form
and we derive a Lagrangian for such equations. In section \ref{two-comp}, we convert our equation to a two-component form and derive a degenerate Lagrangian for this system. In section \ref{Hamilton}, we transform the Lagrangian system into Hamiltonian system using the Dirac's theory of constraints. We obtain the symplectic operator and symplectic two-form, Hamiltonian operator $J_0$ and corresponding Hamiltonian density $H_1$. In section \ref{symmetry}, we present a symmetry condition for one-component form of the symplectic Monge--Amp\`ere equation in terms of first-order linear differential operators $L_{ij(k)}$ and show how the skew-factorized form of the symmetry condition immediately yields Lax pairs and recursion relations. In section \ref{sec-integr}, we develop a general approach for deriving the skew-factorized form of the symmetry condition together with five nontrivial examples. The number of such explicitly integrable examples can be increased by applying permutations of indices together with appropriate permutations of coefficients. In section \ref{recursoper}, we derive a recursion operator $R$ in a $2\times 2$ matrix form for the two-component form of our first equation in section \ref{sec-integr}.
In section \ref{sec-biHam}, by composing the recursion operator with the Hamiltonian operator $J_0$ we obtain the second Hamiltonian operator $J_1=RJ_0$
and the corresponding Hamiltonian density $H_0$ under one constraint on the coefficients of our equation. Thus, we show that our two-component system is a bi-Hamiltonian system integrable in the sense of Magri \cite{magri}. In section \ref{sec-biHam2}, we present the recursion operator, the first and the second Hamiltonian operator, the corresponding Hamiltonian densities and a bi-Hamiltonian representation for further four integrable equations from the section \ref{sec-integr}.

\section{Second-order Lagrangian equations of\\ evolutionary Hirota type}
\setcounter{equation}{0}
\label{Lagrange}

The Fr\'echet derivative operator (linearization) of equation \eqref{F} reads
\begin{eqnarray}
 && D_F = - g D_t^2 + (f_{u_{t1}} - u_{tt}g_{u_{t1}})D_tD_1 + (f_{u_{t2}} - u_{tt}g_{u_{t2}})D_tD_2\nonumber\\
 && \mbox{} + (f_{u_{t3}} - u_{tt}g_{u_{t3}})D_tD_3 + (f_{u_{11}} - u_{tt}g_{u_{11}})D_1^2
  \label{DF}\\
 && \mbox{} + (f_{u_{12}} - u_{tt}g_{u_{12}})D_1D_2 + (f_{u_{13}} - u_{tt}g_{u_{13}})D_1D_3 \nonumber\\
 && \mbox{} + (f_{u_{22}} - u_{tt}g_{u_{22}})D_2^2 + (f_{u_{23}} - u_{tt}g_{u_{23}})D_2D_3 + (f_{u_{33}} - u_{tt}g_{u_{33}})D_3^2 \nonumber
\end{eqnarray}
where $D_i, D_t$ denote operators of total derivatives. The adjoint Fr\'echet derivative operator has the form
\begin{eqnarray}
 && D^*_F = - D_t^2 g + D_tD_1(f_{u_{t1}} - u_{tt}g_{u_{t1}}) + D_tD_2(f_{u_{t2}} - u_{tt}g_{u_{t2}})\nonumber\\
 && \mbox{} + D_tD_3(f_{u_{t3}} - u_{tt}g_{u_{t3}}) + D_1^2(f_{u_{11}} - u_{tt}g_{u_{11}}) \nonumber\\
 && \mbox{} + D_1D_2(f_{u_{12}} - u_{tt}g_{u_{12}}) + D_1D_3(f_{u_{13}} - u_{tt}g_{u_{13}}) \nonumber\\
 && \mbox{} + D_2^2(f_{u_{22}} - u_{tt}g_{u_{22}}) + D_2D_3(f_{u_{23}} - u_{tt}g_{u_{23}}) + D_3^2(f_{u_{33}} - u_{tt}g_{u_{33}}) \nonumber
 \label{DF*}
\end{eqnarray}
According to Helmholtz conditions \cite{olv}, equation \eqref{F} is an Euler-Lagrange equation for a variational problem iff its Fr\'echet derivative is self-adjoint, $D^*_F =D_F$. Equating to zero coefficients of $D_t$, $D_1$, $D_2$, $D_3$ and the term without operators of total derivatives, we obtain five equations on the functions $f$ and $g$. Consider separately the first equation obtained by equating to zero the coefficient of $D_t$
\begin{eqnarray*}
 &&\hspace*{-14pt} - 2D_t[g] + D_1(f_{u_{t1}} - u_{tt}g_{u_{t1}}) + D_2(f_{u_{t2}} - u_{tt}g_{u_{t2}}) + D_3(f_{u_{t3}} - u_{tt}g_{u_{t3}}) = 0.
\end{eqnarray*}
Splitting this equation in $u_{tt1}, u_{tt2}$ and $u_{tt3}$ we obtain $g_{u_{t1}} = 0$, $g_{u_{t2}} = 0$ and $g_{u_{t3}} = 0$, respectively. Using this result, we obtain the five equations mentioned above in the form
\begin{eqnarray}
 && - 2D_t[g] + D_1[f_{u_{t1}}] + D_2[f_{u_{t2}}] + D_3[f_{u_{t3}}] = 0
 \label{I} \\
 && D_t[f_{u_{t1}}] + 2D_1[f_{u_{11}}] - 2u_{tt}D_1[g_{u_{11}}] - 2u_{tt1}g_{u_{11}} + D_2[f_{u_{12}}] - u_{tt}D_2[g_{u_{12}}] \nonumber\\
 &&\mbox{} - u_{tt2}g_{u_{12}}  + D_3[f_{u_{13}}] - u_{tt}D_3[g_{u_{13}}] - u_{tt3}g_{u_{13}} = 0
 \label{II} \\
 && D_t[f_{u_{t2}}] + 2D_2[f_{u_{22}}] - 2u_{tt}D_2[g_{u_{22}}] - 2u_{tt2}g_{u_{22}} + D_1[f_{u_{12}}] - u_{tt}D_1[g_{u_{12}}]\nonumber\\
 &&\mbox{} - u_{tt1}g_{u_{12}}  + D_3[f_{u_{23}}] - u_{tt}D_3[g_{u_{23}}] - u_{tt3}g_{u_{23}} = 0
 \label{III} \\
 && D_t[f_{u_{t3}}] + 2D_3[f_{u_{33}}] - 2u_{tt}D_1[g_{u_{33}}] - 2u_{tt3}g_{u_{33}} + D_1[f_{u_{13}}] - u_{tt}D_1[g_{u_{13}}]\nonumber\\
 && \mbox{} - u_{tt1}g_{u_{13}} + D_2[f_{u_{23}}] - u_{tt}D_2[g_{u_{23}}] - u_{tt2}g_{u_{23}} = 0
 \label{IV} \\
 && D_t^2[g] + D_tD_1[f_{u_{t1}}] + D_tD_2[f_{u_{t2}}] + D_tD_3[f_{u_{t3}}] + D_1^2[f_{u_{11}}-u_{tt}g_{u_{11}}]\nonumber\\
 &&\mbox{} + D_1D_2[f_{u_{12}}-u_{tt}g_{u_{12}}] + D_1D_3[f_{u_{13}}-u_{tt}g_{u_{13}}] + D_2^2[f_{u_{22}}-u_{tt}g_{u_{22}}]\nonumber\\
 && \mbox{} + D_2D_3[f_{u_{23}}-u_{tt}g_{u_{23}}] + D_3^2[f_{u_{33}}-u_{tt}g_{u_{33}}] = 0.\nonumber\\
 \label{V}
\end{eqnarray}
The general solution to these equations for $f$ and $g$ implies the Lagrangian evolutionary Hirota equation \eqref{F} to have the symplectic Monge--Amp\`ere form
\begin{eqnarray}
&& \hspace*{-7pt}\mbox{} F = a_1\{u_{tt}(u_{11}u_{22} - u_{12}^2) - u_{t1}(u_{t1}u_{22} - u_{t2}u_{12}) + u_{t2}(u_{t1}u_{12} - u_{t2}u_{11})\}\nonumber\\
&& \hspace*{-7pt}\mbox{} + a_2\{u_{tt}(u_{11}u_{33} - u_{13}^2) - u_{t1}(u_{t1}u_{33} - u_{t3}u_{13}) + u_{t3}(u_{t1}u_{13} - u_{t3}u_{11})\}\nonumber\\
&& \hspace*{-7pt}\mbox{} + a_3\{u_{tt}(u_{22}u_{33} - u_{23}^2) - u_{t2}(u_{t2}u_{33} - u_{t3}u_{23}) + u_{t3}(u_{t2}u_{23} - u_{t3}u_{22})\}\nonumber\\
&& \hspace*{-7pt}\mbox{} + a_4\{u_{tt}(u_{11}u_{23} - u_{12}u_{13}) - u_{t1}(u_{t1}u_{23} - u_{t2}u_{13}) + u_{t3}(u_{t1}u_{12} - u_{t2}u_{11})\}\nonumber\\
&& \hspace*{-7pt}\mbox{} + a_5\{u_{tt}(u_{12}u_{23} - u_{13}u_{22}) - u_{t1}(u_{t2}u_{23} - u_{t3}u_{22}) + u_{t2}(u_{t2}u_{13} - u_{t3}u_{12})\}\nonumber\\
&& \hspace*{-7pt}\mbox{} + a_6\{u_{tt}(u_{12}u_{33} - u_{13}u_{23}) - u_{t2}(u_{t1}u_{33} - u_{t3}u_{13}) + u_{t3}(u_{t1}u_{23} - u_{t3}u_{12})\}\nonumber\\
&&\hspace*{-7pt}\mbox{} + a_7(u_{tt}u_{11}-u_{t1}^2) + a_8(u_{tt}u_{12}-u_{t1}u_{t2}) + a_9(u_{tt}u_{13} - u_{t1}u_{t3}) \nonumber\\
&&\hspace*{-7pt}\mbox{} + a_{10}(u_{tt}u_{22}-u_{t2}^2) + a_{11}(u_{tt}u_{23}-u_{t2}u_{t3}) + a_{12}(u_{tt}u_{33} - u_{t3}^2) + a_{13}u_{tt} \nonumber\\
&& \hspace*{-7pt}\mbox{} + b_{1}\{u_{t1}(u_{12}u_{23} - u_{13}u_{22}) - u_{t2}(u_{11}u_{23} - u_{12}u_{13}) + u_{t3}(u_{11}u_{22} - u_{12}^2)\}\nonumber\\
&& \hspace*{-7pt}\mbox{} + b_{2}\{u_{t1}(u_{12}u_{33} - u_{13}u_{23}) - u_{t2}(u_{11}u_{33} - u_{13}^2) + u_{t3}(u_{11}u_{23} - u_{12}u_{13})\}\nonumber\\
&& \hspace*{-7pt}\mbox{} + b_{3}\{u_{t1}(u_{22}u_{33} - u_{23}^2) - u_{t2}(u_{12}u_{33} - u_{13}u_{23}) + u_{t3}(u_{12}u_{23} - u_{13}u_{22})\}\nonumber\\
&& \hspace*{-7pt}\mbox{} + b_{4}\{u_{11}(u_{22}u_{33} - u_{23}^2) - u_{12}(u_{12}u_{33} - u_{13}u_{23}) + u_{13}(u_{12}u_{23} - u_{13}u_{22})\}\nonumber\\
&&\hspace*{-7pt}\mbox{} + c_1(u_{t1}u_{12}-u_{t2}u_{11}) + c_2(u_{t1}u_{13} - u_{t3}u_{11}) + c_3(u_{t1}u_{22} - u_{t2}u_{12}) \nonumber\\
&& \hspace*{-7pt}\mbox{} + c_4(u_{t1}u_{23}-u_{t2}u_{13}) + c_5 (u_{t2}u_{23} - u_{t3}u_{22}) + c_6(u_{t1}u_{33} - u_{t3}u_{13})\nonumber\\
&& \hspace*{-7pt}\mbox{} + c_7(u_{t2}u_{33}-u_{t3}u_{23}) + c_8 (u_{t2}u_{13} - u_{t3}u_{12}) + c_{8'} (u_{t1}u_{23} - u_{t3}u_{12})\nonumber\\
&& \hspace*{-7pt}\mbox{} + c_9(u_{11}u_{23} - u_{12}u_{13}) + c_{10}(u_{12}u_{23}-u_{13}u_{22}) + c_{11} (u_{12}u_{33} - u_{13}u_{23})\nonumber\\
&& \hspace*{-7pt}\mbox{} + c_{12}(u_{11}u_{22} - u_{12}^2) + c_{13}(u_{11}u_{33}-u_{13}^2) + c_{14} (u_{22}u_{33} - u_{23}^2)  \nonumber\\
&& \hspace*{-7pt}\mbox{} + c_{15}u_{t1} + c_{16}u_{t2} + c_{17}u_{t3} + c_{18}u_{11} + c_{19}u_{12} + c_{20}u_{13} + c_{21}u_{22} + c_{22}u_{23}\nonumber\\
&& \hspace*{-7pt}\mbox{} + c_{23}u_{33} + c_{24} = 0.
\label{1comp}
\end{eqnarray}
Here we have included the term with the coefficient $c_{8'}$, which is linearly dependent on the terms with coefficients $c_4$ and $c_8$, to make the equation \eqref{1comp} admit the discrete symmetry of permutations of the indices in $u_{ij}$ together with an appropriate permutation of the coefficients.

A Lagrangian for the equation \eqref{1comp} is readily obtained by applying the homotopy formula \cite{olv} for $F = f- u_{tt}g$
\[L[u] = \int\limits_0^1 u\cdot F[\lambda u]\,d\lambda = \int\limits_0^1 u\cdot f[\lambda u]\,d\lambda - \int\limits_0^1 u\cdot(\lambda u_{tt})g[\lambda u]\, d\lambda\]
where $F$ is explicitly given in \eqref{1comp}, with the result
\begin{eqnarray}
 && \hspace*{-20pt} L = \frac{u}{4} \left\langle a_1\{u_{tt}(u_{11}u_{22} - u_{12}^2) - u_{t1}(u_{t1}u_{22} - u_{t2}u_{12}) + u_{t2}(u_{t1}u_{12} - u_{t2}u_{11}) \} \right.  \nonumber\\
 && \hspace*{-20pt}\left. \mbox{} + a_2\{u_{tt}(u_{11}u_{33} - u_{13}^2) - u_{t1}(u_{t1}u_{33} - u_{t3}u_{13}) + u_{t3}(u_{t1}u_{13} - u_{t3}u_{11}) \} \right.  \nonumber\\
 && \hspace*{-20pt}\left. \mbox{} + a_3\{u_{tt}(u_{22}u_{33} - u_{23}^2) - u_{t2}(u_{t2}u_{33} - u_{t3}u_{23}) + u_{t3}(u_{t2}u_{23} - u_{t3}u_{22}) \} \right.  \nonumber\\
 && \hspace*{-20pt}\left. \mbox{} + a_4\{u_{tt}(u_{11}u_{23} - u_{12}u_{13}) - u_{t1}(u_{t1}u_{23} - u_{t2}u_{13}) + u_{t3}(u_{t1}u_{12} - u_{t2}u_{11}) \} \right.  \nonumber\\
 && \hspace*{-20pt}\left. \mbox{} + a_5\{u_{tt}(u_{12}u_{23} - u_{13}u_{22}) - u_{t1}(u_{t2}u_{23} - u_{t3}u_{22}) + u_{t2}(u_{t2}u_{13} - u_{t3}u_{12}) \} \right.  \nonumber\\
 && \hspace*{-20pt}\left. \mbox{} + a_6\{u_{tt}(u_{12}u_{33} - u_{13}u_{23}) - u_{t1}(u_{t2}u_{33} - u_{t3}u_{23}) + u_{t3}(u_{t2}u_{13} - u_{t3}u_{12}) \} \right.  \nonumber\\
 && \hspace*{-20pt}\left. \mbox{} +  b_{1}\{u_{t1}(u_{12}u_{23} - u_{13}u_{22}) - u_{t2}(u_{11}u_{23} - u_{12}u_{13}) + u_{t3}(u_{11}u_{22} - u_{12}^2)\}\right. \nonumber\\
 && \hspace*{-20pt} \left.\mbox{} + b_{2}\{u_{t1}(u_{12}u_{33} - u_{13}u_{23}) - u_{t2}(u_{11}u_{33} - u_{13}^2) + u_{t3}(u_{11}u_{23} - u_{12}u_{13})\} \right. \nonumber\\
 && \hspace*{-20pt} \left.\mbox{} + b_{3}\{u_{t1}(u_{22}u_{33} - u_{23}^2) - u_{t2}(u_{12}u_{33} - u_{13}u_{23}) + u_{t3}(u_{12}u_{23} - u_{13}u_{22})\}\right. \nonumber\\
 && \hspace*{-20pt} \left.\mbox{} + b_{4}\{u_{11}(u_{22}u_{33} - u_{23}^2) - u_{12}(u_{12}u_{33} - u_{13}u_{23}) + u_{13}(u_{12}u_{23} - u_{13}u_{22})\}
 \right\rangle \nonumber\\
 && \hspace*{-20pt} \mbox{} + \frac{u}{3}\bigl\{a_7(u_{tt}u_{11}-u_{t1}^2) + a_8(u_{tt}u_{12}-u_{t1}u_{t2}) + a_9(u_{tt}u_{13}-u_{t1}u_{t3}) \nonumber\\
  && \hspace*{-20pt}\mbox{} + a_{10}(u_{tt}u_{22}-u_{t2}^2) + a_{11}(u_{tt}u_{23}-u_{t2}u_{t3}) + a_{12}(u_{tt}u_{33}-u_{t3}^2) \nonumber\\
 && \hspace*{-20pt} \mbox{} +  c_1(u_{t1}u_{12}-u_{t2}u_{11}) + c_2(u_{t1}u_{13}-u_{t3}u_{11}) + c_3(u_{t1}u_{22}-u_{t2}u_{12})\nonumber\\
 && \hspace*{-20pt} \mbox{} + c_4(u_{t1}u_{23}-u_{t2}u_{13}) + c_5 (u_{t2}u_{23}-u_{t3}u_{22}) + c_6(u_{t1}u_{33}-u_{t3}u_{13})\nonumber\\
 && \hspace*{-20pt} \mbox{} + c_7(u_{t2}u_{33}-u_{t3}u_{23}) + c_8 (u_{t2}u_{13}-u_{t3}u_{12})  + c_{8'} (u_{t1}u_{23}-u_{t3}u_{12})\nonumber\\
 && \hspace*{-20pt} \mbox{} + c_9(u_{11}u_{23}-u_{12}u_{13}) + c_{10}(u_{12}u_{23}-u_{13}u_{22}) + c_{11} (u_{12}u_{33}-u_{13}u_{23})\nonumber\\
 && \hspace*{-20pt} \mbox{} + c_{12}(u_{11}u_{22}-u_{12}^2) + c_{13}(u_{11}u_{33}-u_{13}^2) + c_{14} (u_{22}u_{33}-u_{23}^2)\bigr\} \nonumber\\
 && \hspace*{-20pt} \mbox{} + \frac{u}{2}(a_{13}u_{tt} + c_{15}u_{t1} + c_{16}u_{t2} + c_{17}u_{t3} + c_{18}u_{11} + c_{19}u_{12} + c_{20}u_{13}\nonumber\\
 && \hspace*{-20pt} \mbox{} + c_{21}u_{22} + c_{22}u_{23} + c_{23}u_{33}) + c_{24}u.
  \label{L}
\end{eqnarray}

\section{Two-component form}
\setcounter{equation}{0}
\label{two-comp}

Introducing the second component $v=u_t$ and solving equation \eqref{1comp} with respect to $v_t=u_{tt}$, we convert \eqref{1comp} into the evolutionary two-component system
\begin{eqnarray}
&& \hspace*{-23pt} u_t = v, \nonumber\\
&& \hspace*{-23pt} v_{t} = \frac{1}{\Delta} \Bigl\langle a_1(v_1^2u_{22} + v_2^2u_{11} - 2v_1v_2u_{12}) + a_2(v_1^2u_{33} + v_3^2u_{11} - 2v_1v_3u_{13}) \nonumber\\
&& \hspace*{-25pt} \mbox{} + a_3(v_2^2u_{33} + v_3^2u_{22} - 2v_2v_3u_{23}) + a_4\{v_1(v_1u_{23} - v_2u_{13}) - v_3(v_1u_{12} - v_2u_{11})\} \nonumber\\
&& \hspace*{-23pt} \mbox{} + a_5\{v_1(v_2u_{23} - v_3u_{22}) - v_2(v_2u_{13} - v_3u_{12})\}  \nonumber\\
&& \hspace*{-23pt} \mbox{} + a_6\{v_2(v_1u_{33} - v_3u_{13}) - v_3(v_1u_{23} - v_3u_{12})\} + a_7v_1^2 + a_8v_1v_2 + a_9v_1v_3
\nonumber\\
&& \hspace*{-23pt} \mbox{} + a_{10}v_2^2 + a_{11}v_2v_3 + a_{12}v_3^2  \nonumber\\
&& \hspace*{-23pt} \mbox{} - b_{1}\{v_{1}(u_{12}u_{23} - u_{13}u_{22}) - v_{2}(u_{11}u_{23} - u_{12}u_{13}) + v_{3}(u_{11}u_{22} - u_{12}^2)\}\nonumber\\
&& \hspace*{-23pt}\mbox{} - b_{2}\{v_{1}(u_{12}u_{33} - u_{13}u_{23}) - v_{2}(u_{11}u_{33} - u_{13}^2) + v_{3}(u_{11}u_{23} - u_{12}u_{13})\}\nonumber\\
&& \hspace*{-23pt}\mbox{} - b_{3}\{v_{1}(u_{22}u_{33} - u_{23}^2) - v_{2}(u_{12}u_{33} - u_{13}u_{23}) + v_{3}(u_{12}u_{23} - u_{13}u_{22})\}\nonumber\\
&& \hspace*{-23pt}\mbox{} - b_{4}\{u_{11}(u_{22}u_{33} - u_{23}^2) - u_{12}(u_{12}u_{33} - u_{13}u_{23}) + u_{13}(u_{12}u_{23} - u_{13}u_{22})\}\nonumber\\
&&\hspace*{-23pt}\mbox{} - c_1(v_{1}u_{12}-v_{2}u_{11}) - c_2(v_{1}u_{13}-v_{3}u_{11}) - c_3(v_{1}u_{22} - v_{2}u_{12}) \nonumber\\
&& \hspace*{-23pt}\mbox{} - c_4(v_{1}u_{23}-v_{2}u_{13}) - c_5 (v_{2}u_{23}-v_{3}u_{22}) - c_6(v_{1}u_{33}-v_{3}u_{13})\nonumber\\
&& \hspace*{-23pt}\mbox{} - c_7(v_{2}u_{33}-v_{3}u_{23}) - c_8 (v_{2}u_{13}-v_{3}u_{12})  - c_{8'} (v_{1}u_{23}-v_{3}u_{12})\nonumber\\
&& \hspace*{-23pt}\mbox{} - c_9(u_{11}u_{23}-u_{12}u_{13}) - c_{10}(u_{12}u_{23}-u_{13}u_{22}) - c_{11} (u_{12}u_{33}-u_{13}u_{23})\nonumber\\
&& \hspace*{-23pt}\mbox{} - c_{12}(u_{11}u_{22}-u_{12}^2) - c_{13}(u_{11}u_{33}-u_{13}^2) - c_{14} (u_{22}u_{33}-u_{23}^2) \nonumber\\
&& \hspace*{-23pt}\mbox{} - c_{15}v_{1} - c_{16}v_{2} - c_{17}v_{3} - c_{18}u_{11} - c_{19}u_{12} - c_{20}u_{13}\nonumber\\
&& \hspace*{-23pt}\mbox{}- c_{21}u_{22} - c_{22}u_{23} - c_{23}u_{33} - c_{24}\Bigr\rangle\nonumber\\
&& \hspace*{-23pt}\mbox{}\equiv \frac{1}{\Delta}\left(\sum_{i=1}^{12}a_iq^{(ai)} + \sum_{i=1}^4b_iq^{(bi)} + \sum_{i=1}^{\hspace*{4pt}24\hspace*{4pt}\prime}c_iq^{(i)}\right)\equiv\frac{q}{\Delta}
\label{2comp}
\end{eqnarray}
where the last sum includes also $i=8'$ and
\begin{eqnarray}
&& \Delta = a_1(u_{11}u_{22} - u_{12}^2) + a_2(u_{11}u_{33} - u_{13}^2) + a_3(u_{22}u_{33} - u_{23}^2) \nonumber\\
&& \mbox{} + a_4(u_{11}u_{23} - u_{12}u_{13}) + a_5(u_{12}u_{23} - u_{13}u_{22}) + a_6(u_{12}u_{33} - u_{13}u_{23}) \nonumber\\
&& \mbox{} + a_7u_{11} + a_8u_{12} + a_9u_{13} + a_{10}u_{22} + a_{11}u_{23} + a_{12}u_{33} + a_{13}.
 \label{Delt}
\end{eqnarray}

The Lagrangian for the system \eqref{2comp} is obtained by a suitable modification of the Lagrangian \eqref{L} of the one-component equation \eqref{1comp}, skipping some total derivative terms
\begin{eqnarray}
 && \hspace*{-14.2pt} L = \left(u_tv - \frac{1}{2}v^2\right)\{a_1(u_{11}u_{22}-u_{12}^2) + a_2(u_{11}u_{33}-u_{13}^2) + a_3(u_{22}u_{33}-u_{23}^2) \nonumber\\
 && \hspace*{-14.2pt} \mbox{} + a_4(u_{11}u_{23}-u_{12}u_{13}) + a_5(u_{12}u_{23}-u_{13}u_{22}) + a_6(u_{12}u_{33}-u_{13}u_{23})  \nonumber\\
 && \hspace*{-14.2pt} \mbox{} + a_7u_{11} + a_8u_{12} + a_9u_{13} + a_{10}u_{22} + a_{11}u_{23} + a_{12}u_{33} + a_{13}\} \nonumber\\
 && \hspace*{-14.2pt} \mbox{} + \frac{u_t}{4} \left\langle b_{1}\{u_{1}(u_{12}u_{23} - u_{13}u_{22}) - u_{2}(u_{11}u_{23} - u_{12}u_{13}) + u_{3}(u_{11}u_{22} - u_{12}^2)\}\right. \nonumber\\
 && \hspace*{-14.2pt} \left.\mbox{} + b_{2}\{u_{1}(u_{12}u_{33} - u_{13}u_{23}) - u_{2}(u_{11}u_{33} - u_{13}^2) + u_{3}(u_{11}u_{23} - u_{12}u_{13})\} \right. \nonumber\\
 && \hspace*{-14.2pt} \left.\mbox{} + b_{3}\{u_{1}(u_{22}u_{33} - u_{23}^2) - u_{2}(u_{12}u_{33} - u_{13}u_{23}) + u_{3}(u_{12}u_{23} - u_{13}u_{22})\} \right\rangle \nonumber\\
 && \hspace*{-14.2pt} \mbox{} - b_{4}\frac{u}{4}\{u_{11}(u_{22}u_{33} - u_{23}^2) - u_{12}(u_{12}u_{33} - u_{13}u_{23}) + u_{13}(u_{12}u_{23} - u_{13}u_{22})\}
  \nonumber\\
 && \hspace*{-14.2pt} \mbox{} + \frac{u_t}{3}\bigl\{c_1(u_{1}u_{12}-u_{2}u_{11}) + c_2(u_{1}u_{13}-u_{3}u_{11}) + c_3(u_{1}u_{22}-u_{2}u_{12})\nonumber\\
 && \hspace*{-14.2pt} \mbox{} + c_4(u_{1}u_{23}-u_{2}u_{13}) + c_5 (u_{2}u_{23}-u_{3}u_{22}) + c_6(u_{1}u_{33}-u_{3}u_{13})\nonumber\\
 && \hspace*{-14.2pt} \mbox{} + c_7(u_{2}u_{33}-u_{3}u_{23}) + c_8 (u_{2}u_{13}-u_{3}u_{12}) + c_{8'} (u_{1}u_{23}-u_{3}u_{12})\}\nonumber\\
 && \hspace*{-14.2pt} \mbox{}- \frac{u}{3} \bigl\{c_9(u_{11}u_{23}-u_{12}u_{13}) + c_{10}(u_{12}u_{23}-u_{13}u_{22}) + c_{11}(u_{12}u_{33}-u_{13}u_{23})\nonumber\\
 && \hspace*{-14.2pt} \mbox{} + c_{12}(u_{11}u_{22}-u_{12}^2) + c_{13}(u_{11}u_{33}-u_{13}^2) + c_{14} (u_{22}u_{33}-u_{23}^2)\bigr\} \nonumber\\
 && \hspace*{-14.2pt} \mbox{} + \frac{u_t}{2} (c_{15}u_{1} + c_{16}u_{2} + c_{17}u_{3})\nonumber\\
 && \hspace*{-14.2pt} \mbox{} - \frac{u}{2} (c_{18}u_{11} + c_{19}u_{12} + c_{20}u_{13} + c_{21}u_{22} + c_{22}u_{23} + c_{23}u_{33}) - c_{24}u
  \label{L2}
\end{eqnarray}
where we have changed the overall sign of $L$.

\section{Hamiltonian representation}
\setcounter{equation}{0}
\label{Hamilton}

To transform from Lagrangian to Hamiltonian description, we define the canonical momenta
\begin{eqnarray}
 && \hspace*{-12pt} \pi_u = \frac{\partial L}{\partial u_t} = v\{a_1(u_{11}u_{22}-u_{12}^2) + a_2(u_{11}u_{33}-u_{13}^2) + a_3(u_{22}u_{33}-u_{23}^2) \nonumber\\
 && \hspace*{-12pt} \mbox{} + a_4(u_{11}u_{23}-u_{12}u_{13}) + a_5(u_{12}u_{23}-u_{13}u_{22}) + a_6(u_{12}u_{33}-u_{13}u_{23})  \nonumber\\
 && \hspace*{-12pt} \mbox{} + a_7u_{11} + a_8u_{12} + a_9u_{13} + a_{10}u_{22} + a_{11}u_{23} + a_{12}u_{33} + a_{13}\} \nonumber\\
 && \hspace*{-12pt} \mbox{} + \frac{1}{4} \left\langle b_{1}\{u_{1}(u_{12}u_{23} - u_{13}u_{22}) - u_{2}(u_{11}u_{23} - u_{12}u_{13}) + u_{3}(u_{11}u_{22} - u_{12}^2)\}\right.  \nonumber\\
 && \hspace*{-12pt} \left.\mbox{} + b_{2}\{u_{1}(u_{12}u_{33} - u_{13}u_{23}) - u_{2}(u_{11}u_{33} - u_{13}^2) + u_{3}(u_{11}u_{23} - u_{12}u_{13})\} \right. \nonumber\\
 && \hspace*{-12pt} \left.\mbox{} + b_{3}\{u_{1}(u_{22}u_{33} - u_{23}^2) - u_{2}(u_{12}u_{33} - u_{13}u_{23}) + u_{3}(u_{12}u_{23} - u_{13}u_{22})\} \right\rangle \nonumber\\
  && \hspace*{-12pt} \mbox{} + \frac{1}{3}\bigl\{c_1(u_{1}u_{12}-u_{2}u_{11}) + c_2(u_{1}u_{13}-u_{3}u_{11}) + c_3(u_{1}u_{22}-u_{2}u_{12})\nonumber\\
 && \hspace*{-12pt} \mbox{} + c_4(u_{1}u_{23}-u_{2}u_{13}) + c_5 (u_{2}u_{23}-u_{3}u_{22}) + c_6(u_{1}u_{33}-u_{3}u_{13})\nonumber\\
 && \hspace*{-12pt} \mbox{} + c_7(u_{2}u_{33}-u_{3}u_{23}) + c_8 (u_{2}u_{13}-u_{3}u_{12}) + c_{8'} (u_{1}u_{23}-u_{3}u_{12})\bigr\}\nonumber\\
 && \hspace*{-12pt} \mbox{} + \frac{1}{2} (c_{15}u_{1} + c_{16}u_{2} + c_{17}u_{3}),\quad \pi_v = \frac{\partial L}{\partial v_t} = 0
  \label{mom}
\end{eqnarray}
which satisfy canonical Poisson brackets $[u^i(z),\pi^k(z')] = \delta^{ik} \delta(z-z')$,
where $u^1=u$, $u^2=v$, $\pi^1=\pi_u$, $\pi^2=\pi_v$ and $z=(z_1,z_2,z_3)$.
The Lagrangian \eqref{L2} is degenerate because the momenta cannot be inverted for the velocities. Therefore, following Dirac's theory of constraints \cite{dirac}, we impose \eqref{mom} as constraints $\Phi_u = 0$, $\Phi_v = 0$ where
\begin{eqnarray}
 && \Phi_u =  \pi_u - v\{a_1(u_{11}u_{22}-u_{12}^2) + a_2(u_{11}u_{33}-u_{13}^2) + a_3(u_{22}u_{33}-u_{23}^2) \nonumber\\
 && \hspace*{-12pt} \mbox{} + a_4(u_{11}u_{23}-u_{12}u_{13}) + a_5(u_{12}u_{23}-u_{13}u_{22}) + a_6(u_{12}u_{33}-u_{13}u_{23})  \nonumber\\
 && \hspace*{-12pt} \mbox{} + a_7u_{11} + a_8u_{12} + a_9u_{13} + a_{10}u_{22} + a_{11}u_{23} + a_{12}u_{33} + a_{13}\} \nonumber\\
 && \hspace*{-12pt} \mbox{} - \frac{1}{4} \left\langle b_{1}\{u_{1}(u_{12}u_{23} - u_{13}u_{22}) - u_{2}(u_{11}u_{23} - u_{12}u_{13}) + u_{3}(u_{11}u_{22} - u_{12}^2)\}\right.  \nonumber\\
 && \hspace*{-12pt} \left.\mbox{} + b_{2}\{u_{1}(u_{12}u_{33} - u_{13}u_{23}) - u_{2}(u_{11}u_{33} - u_{13}^2) + u_{3}(u_{11}u_{23} - u_{12}u_{13})\} \right. \nonumber\\
 && \hspace*{-12pt} \left.\mbox{} + b_{3}\{u_{1}(u_{22}u_{33} - u_{23}^2) - u_{2}(u_{12}u_{33} - u_{13}u_{23}) + u_{3}(u_{12}u_{23} - u_{13}u_{22})\} \right\rangle \nonumber\\
  && \hspace*{-12pt} \mbox{} - \frac{1}{3}\bigl\{c_1(u_{1}u_{12}-u_{2}u_{11}) + c_2(u_{1}u_{13}-u_{3}u_{11}) + c_3(u_{1}u_{22}-u_{2}u_{12})\nonumber\\
 && \hspace*{-12pt} \mbox{} + c_4(u_{1}u_{23}-u_{2}u_{13}) + c_5 (u_{2}u_{23}-u_{3}u_{22}) + c_6(u_{1}u_{33}-u_{3}u_{13})
     \label{dirac} \\
 && \hspace*{-12pt} \mbox{} + c_7(u_{2}u_{33}-u_{3}u_{23}) + c_8 (u_{2}u_{13}-u_{3}u_{12})  + c_{8'} (u_{1}u_{23}-u_{3}u_{12})\bigr\}\nonumber\\
 && \hspace*{-12pt} \mbox{} - \frac{1}{2} (c_{15}u_{1} + c_{16}u_{2} + c_{17}u_{3}) \nonumber\\
 && \hspace*{-12pt} \Phi_v = \pi_v
\end{eqnarray}
and calculate Poisson brackets for the constraints
\begin{eqnarray}
 && K_{11} = [\Phi_u(z),\Phi_{u'}(z')], \quad K_{12} = [\Phi_u(z),\Phi_{v'}(z')] \nonumber\\
 && K_{21} = [\Phi_v(z),\Phi_{u'}(z')],\quad K_{22} = [\Phi_v(z),\Phi_{v'}(z')].
 \label{constr}
\end{eqnarray}
We obtain the following matrix of Poisson brackets%, which for convenience we multiply by the overall factor $(-1)$
\begin{equation}
  K = \left(
  \begin{array}{cc}
   K_{11} & K_{12} \\
 - K_{12} &  0
  \end{array}
  \right)
\label{K}
\end{equation}
where
\begin{equation}
  K_{11} = \sum_{i=1}^{13}a_iK_{11}^{(ai)} + \sum_{i=1}^3b_iK_{11}^{(bi)} + \sum_{i=1}^{8'}c_iK_{11}^{(i)} - \sum_{i=1}^3c_{i+14}D_i,\quad K_{12}=\sum_{i=1}^{13}a_iK_{12}^{(i)}
 \label{Kij}
\end{equation}
with the following definitions
\begin{eqnarray}
 && K_{11}^{(a1)} = 2(v_1u_{22} - v_2u_{12})D_1 + 2(v_2u_{11} - v_1u_{12})D_2  + v_{11}u_{22} + v_{22}u_{11} \nonumber\\
 && \mbox{} - 2v_{12}u_{12}, \quad K_{12}^{(1)} = - (u_{11}u_{22} - u_{12}^2),\quad  K_{11}^{(a2)} = 2(v_1u_{33} - v_3u_{13})D_1 \nonumber\\
 &&\mbox{} + 2(v_3u_{11} - v_1u_{13})D_3 + v_{11}u_{33} + v_{33}u_{11} - 2v_{13}u_{13}, \nonumber\\
 && \mbox{} \quad K_{12}^{(2)} = - (u_{11}u_{33} - u_{13}^2),\quad K_{11}^{(a3)} = 2(v_2u_{33} - v_3u_{23})D_2 \nonumber
 %  \label{KijP1}
 \\
 &&\hspace*{-31pt}\mbox{} + 2(v_3u_{22} - v_2u_{23})D_3  + v_{22}u_{33} + v_{33}u_{22} - 2v_{23}u_{23},\quad K_{12}^{(3)} = - (u_{22}u_{33} - u_{23}^2). \nonumber \\
 &&\hspace*{-31pt} K_{11}^{(a4)} = (2v_1u_{23} - v_2u_{13} - v_3u_{12})D_1 + (v_3u_{11} - v_1u_{13})D_2 + (v_2u_{11} - v_1u_{12})D_3 \nonumber\\
 && \mbox{}  + v_{11}u_{23} + v_{23}u_{11} - v_{12}u_{13} - v_{13}u_{12},\quad K_{12}^{(4)} = - (u_{11}u_{23} - u_{12}u_{13})  \nonumber\\
 &&\hspace*{-31pt} K_{11}^{(a5)} = (v_2u_{23} - v_3u_{22})D_1 + (v_1u_{23} - 2v_2u_{13} + v_3u_{12})D_2 + (v_2u_{12} - v_1u_{22})D_3 \nonumber\\
 && \mbox{} + v_{12}u_{23} + v_{23}u_{12} - v_{13}u_{22} - v_{22}u_{13},\quad K_{12}^{(5)} = - (u_{12}u_{23} - u_{13}u_{22}) \nonumber\\
 &&\hspace*{-41pt} \quad K_{12}^{(6)} = - (u_{12}u_{33} - u_{13}u_{23}),\quad K_{11}^{(a6)} = (v_2u_{33} - v_3u_{23})D_1 + (v_1u_{33} - v_3u_{13})D_2   \nonumber\\
 && \mbox{} + (2v_3u_{12} - v_1u_{23} - v_2u_{13})D_3 + v_{12}u_{33} + v_{33}u_{12} - v_{13}u_{23} - v_{23}u_{13}  \nonumber\\
 && K_{11}^{(a7)} = 2v_1D_1 + v_{11},\; K_{12}^{(7)} = - u_{11},\quad K_{11}^{(a8)} = v_2D_1 + v_1D_2 + v_{12},\nonumber\\
 && K_{12}^{(8)} = - u_{12},\quad K_{11}^{(a9)} = v_3D_1 + v_1D_3 + v_{13},\; K_{12}^{(9)} = - u_{13}    \nonumber\\
 && K_{11}^{(a10)} = 2v_2D_2 + v_{22},\; K_{12}^{(10)} = - u_{22},\quad K_{11}^{(a11)} = v_3D_2 + v_2D_3 + v_{23}\nonumber\\
 && K_{12}^{(11)} = - u_{23},\quad K_{11}^{(a12)} = 2v_3D_3 + v_{33},\; K_{12}^{(12)} = - u_{33},\quad K_{11}^{(a13)} = 0\nonumber\\
 && K_{12}^{(13)} = - 1.
 \label{KijP}
\end{eqnarray}
 \begin{eqnarray}
 &&\hspace*{-10.1pt} K_{11}^{(b1)} = (u_{13}u_{22} - u_{12}u_{23})D_1 + (u_{11}u_{23} - u_{12}u_{13})D_2 - (u_{11}u_{22} - u_{12}^2)D_3\nonumber\\
 &&\hspace*{-10.1pt} K_{11}^{(b2)} = (u_{13}u_{23} - u_{12}u_{33})D_1 + (u_{11}u_{33} - u_{13}^2)D_2 - (u_{11}u_{23} - u_{12}u_{13})D_3\nonumber\\
 &&\hspace*{-10.1pt} K_{11}^{(b3)} = - (u_{22}u_{33} - u_{23}^2)D_1 + (u_{12}u_{33} - u_{13}u_{23})D_2 - (u_{12}u_{23} - u_{13}u_{22})D_3\nonumber\\
 &&\hspace*{-10.1pt} K_{11}^{(1)} = u_{11}D_2 - u_{12}D_1,\; K_{11}^{(2)} = u_{11}D_3 - u_{13}D_1,\; K_{11}^{(3)} = u_{12}D_2 - u_{22}D_1\nonumber\\
 &&\hspace*{-10.1pt} K_{11}^{(4)} = u_{13}D_2 - u_{23}D_1,\; K_{11}^{(5)} = u_{22}D_3 - u_{23}D_2,\;  K_{11}^{(6)} = u_{13}D_3 - u_{33}D_1\nonumber\\
 &&\hspace*{-10.1pt} K_{11}^{(7)} = u_{23}D_3 - u_{33}D_2,\;  K_{11}^{(8)} = u_{12}D_3 - u_{13}D_2,\; K_{11}^{({8'})} = u_{12}D_3 - u_{23}D_1\nonumber\\
 &&\hspace*{-10.1pt} K_{11}^{(15)} = - D_1, \; K_{11}^{(16)} = - D_2,\; K_{11}^{(17)} = - D_3
 \label{KijP3}
\end{eqnarray}
with all other components of $K_{11}$ vanishing.
The components of $K_{11}$ can be presented in a manifestly skew symmetric form, so that $K$ is skew symmetric.

The Hamiltonian operator is an inverse to the symplectic operator
  \begin{equation}
    J_0 = K^{-1} = \left(
    \begin{array}{cc}
   0 & -K_{12}^{-1}\\
   K_{12}^{-1} & K_{12}^{-1}K_{11}K_{12}^{-1}
      \end{array}
    \right).
\label{hamilton1}
\end{equation}

Operator $J_0$ is Hamiltonian if and only if its inverse $K$ is symplectic \cite{ff}, which means for skew symmetric $K$ that the volume integral of $\omega = (1/2)du^i\wedge K_{ij}du^j$ should be a symplectic form, i.e. at appropriate boundary conditions $d\omega = 0$ modulo total divergence. Here summations over $i,j$ run from 1 to 2 and $u^1 = u,\; u^2 = v$, so that
\begin{eqnarray}
 && \omega = \sum_{i=1}^{13}a_i\omega^a_i + \sum_{i=1}^3b_i\omega^b_i + \sum_{i=1}^8c_i\omega_i + \sum_{i=1}^3c_{i+14}\,\omega_{i+14},\nonumber\\
 && \omega^a_i = \frac{1}{2}du\wedge K_{11}^{(ai)}du + du\wedge K_{12}^{(i)}dv,\quad \omega^b_i = \frac{1}{2}du\wedge K_{11}^{(bi)}du\nonumber\\
 && \omega_i = \frac{1}{2}du\wedge K_{11}^{(i)}du
 \label{omegaK}
\end{eqnarray}
where $K_{12}^{(bi)}=0$, $K_{12}^{(i)}=0$.
Using \eqref{KijP} and \eqref{KijP3} for $K_{11}^{(ai)}$, $K_{11}^{(bi)}$, $K_{11}^{(i)}$
and $K_{12}^{(i)}$ in \eqref{omegaK}, we obtain
\begin{eqnarray}
 && \omega^a_1 = (v_1u_{22} - v_2u_{12})du\wedge du_1 + (v_2u_{11} - v_1u_{12})du\wedge du_2 \nonumber\\
 && \mbox{} - (u_{11}u_{22} - u_{12}^2)du\wedge dv \nonumber\\
 && \omega^a_2 = (v_1u_{33} - v_3u_{13})du\wedge du_1 + (v_3u_{11} - v_1u_{13})du\wedge du_3 \nonumber\\
 && \mbox{} - (u_{11}u_{33} - u_{13}^2)du\wedge dv \nonumber\\
 && \omega^a_3 = (v_2u_{33} - v_3u_{23})du\wedge du_2 + (v_3u_{22} - v_2u_{23})du\wedge du_3 \nonumber\\
 && \mbox{} - (u_{22}u_{33} - u_{23}^2)du\wedge dv \nonumber\\
 && \omega^a_4 = \frac{1}{2}\{(2v_1u_{23} - v_2u_{13} - v_3u_{12})du\wedge du_1 + (v_3u_{11} - v_1u_{13})du\wedge du_2 \nonumber\\
 && \mbox{} + (v_2u_{11} - v_1u_{12})du\wedge du_3\} - (u_{11}u_{23} - u_{12}u_{13})du\wedge dv \nonumber\\
 && \omega^a_5 = \frac{1}{2}\{(- 2v_2u_{13} + v_1u_{23} + v_3u_{12})du\wedge du_2 + (v_2u_{12} - v_1u_{22})du\wedge du_3 \nonumber\\
 && \mbox{} + (v_2u_{23} - v_3u_{22})du\wedge du_1\} - (u_{12}u_{23} - u_{13}u_{22})du\wedge dv \nonumber\\
 && \omega^a_6 = \frac{1}{2}\{(2v_3u_{12} - v_1u_{23} - v_2u_{13})du\wedge du_3 + (v_1u_{33} - v_3u_{13})du\wedge du_2 \nonumber\\
 && \mbox{} + (v_2u_{33} - v_3u_{23})du\wedge du_1\} - (u_{12}u_{33} - u_{13}u_{23})du\wedge dv \nonumber\\
 && \omega^a_7 = v_1du\wedge du_1 - u_{11}du\wedge dv,\quad \omega^a_8 = \frac{1}{2}(v_1du\wedge du_2 + v_2du\wedge du_1) \nonumber\\
 && \mbox{}  - u_{12}du\wedge dv,\quad \omega^a_9 = \frac{1}{2}(v_1du\wedge du_3 + v_3du\wedge du_1) - u_{13}du\wedge dv \nonumber\\
 && \omega^a_{10} = v_2du\wedge du_2 - u_{22}du\wedge dv,\quad \omega^a_{11} = \frac{1}{2}(v_3du\wedge du_2 + v_2du\wedge du_3) \nonumber\\
 && \mbox{} - u_{23}du\wedge dv,\quad \omega^a_{12} = v_3du\wedge du_3 - u_{33}du\wedge dv,\quad \omega^a_{13} = du\wedge dv  \nonumber\\
 && \omega^b_{1} = \frac{1}{2}\{(u_{13}u_{22} - u_{12}u_{23})du\wedge du_1 + (u_{11}u_{23} - u_{12}u_{13})du\wedge du_2 \nonumber\\
 && \mbox{} - (u_{11}u_{22} - u_{12}^2)du\wedge du_3\},\quad \omega^b_{2} = \frac{1}{2}\{(u_{13}u_{23} - u_{12}u_{33})du\wedge du_1 \nonumber\\
 && \mbox{} + (u_{11}u_{33} - u_{13}^2)du\wedge du_2 - (u_{11}u_{23} - u_{12}u_{13})du\wedge du_3\} \nonumber\\
 && \omega^b_{3} = \frac{1}{2}\{(u_{23}^2 - u_{22}u_{33})du\wedge du_1 + (u_{12}u_{33} - u_{13}u_{23})du\wedge du_2 \nonumber\\
 && \mbox{} - (u_{12}u_{23} - u_{13}u_{22})du\wedge du_3\},\quad \omega_{1} = \frac{1}{2}(u_{11}du\wedge du_2 - u_{12}du\wedge du_1) \nonumber
 \end{eqnarray}
\begin{eqnarray}
 &&\hspace*{-15pt} \omega_{2} = \frac{1}{2}(u_{11}du\wedge du_3 - u_{13}du\wedge du_1),\quad \omega_{3} = \frac{1}{2}(u_{12}du\wedge du_2 - u_{22}du\wedge du_1) \nonumber\\
 &&\hspace*{-15pt} \omega_{4} = \frac{1}{2}(u_{13}du\wedge du_2 - u_{23}du\wedge du_1),\quad \omega_{5} = \frac{1}{2}(u_{22}du\wedge du_3 - u_{23}du\wedge du_2) \nonumber\\
 &&\hspace*{-15pt} \omega_{6} = \frac{1}{2}(u_{13}du\wedge du_3 - u_{33}du\wedge du_1),\quad \omega_{7} = \frac{1}{2}(u_{23}du\wedge du_3 - u_{33}du\wedge du_2) \nonumber\\
 &&\hspace*{-15pt} \omega_{8} = \frac{1}{2}(u_{12}du\wedge du_3 - u_{13}du\wedge du_2),\quad \omega_{{8'}} = \frac{1}{2}(u_{12}du\wedge du_3 - u_{23}du\wedge du_1)\nonumber\\
 &&\hspace*{-15pt} \omega_{15} = -\frac{1}{2}du\wedge du_1, \quad \omega_{16} = -\frac{1}{2}du\wedge du_2,\quad \omega_{17} = -\frac{1}{2}du\wedge du_3.
 \label{omega}
\end{eqnarray}
Taking exterior derivatives of \eqref{omega} and skipping total divergence terms, we have checked that $d\omega=0$ modulo total divergence which proves that operator $K$ is symplectic because the closedness condition for $\omega$ is equivalent to the Jacobi identity for $J_0$ \cite{ff}. Hence, $J_0$ defined in \eqref{hamilton1} is indeed a Hamiltonian operator.

Hamiltonian form of this system is
\begin{equation}
\left(
\begin{array}{c}
 u_t\\
 v_t
\end{array}
\right) = J_0
\left(
\begin{array}{c}
 \delta_u H_1\\
 \delta_v H_1
\end{array}
\right)
  \label{Hform1}
  \end{equation}
where we still need to determine the corresponding Hamiltonian density $H_1$. We apply the formula $H_1 = \pi_u u_t + \pi_v v_t - L$, where $\pi_v=0$, with the final result
\begin{eqnarray}
 &&\hspace*{-8.7pt} H_1 = - \frac{v^2}{2}\sum_{i=1}^{13}a_iK_{12}^{(i)} \nonumber\\
 &&\hspace*{-8.7pt}\mbox{} + b_4\frac{u}{4}\{u_{11}(u_{22}u_{33} - u_{23}^2) - u_{12}(u_{12}u_{33} - u_{13}u_{23}) + u_{13}(u_{12}u_{23} - u_{13}u_{22})\}          \nonumber\\
 &&\hspace*{-8.7pt}\mbox{} + \frac{u}{3}\{c_9(u_{11}u_{23} - u_{12}u_{13}) + c_{10}(u_{12}u_{23} - u_{13}u_{22}) + c_{11}(u_{12}u_{33} - u_{13}u_{23})\nonumber\\
 &&\hspace*{-8.7pt}\mbox{} + c_{12}(u_{11}u_{22} - u_{12}^2) + c_{13}(u_{11}u_{33} - u_{13}^2) + c_{14}(u_{22}u_{33} - u_{23}^2)\} \nonumber\\
 &&\hspace*{-8.7pt}\mbox{} + \frac{u}{2} (c_{18}u_{11} + c_{19}u_{12} + c_{20}u_{13} + c_{21}u_{22} + c_{22}u_{23} + c_{23}u_{33}) + c_{24}u.
\label{H1}
\end{eqnarray}
We can write the Hamiltonian density in \eqref{H1} in the following short-hand notation
\begin{equation}
 H_1 = \sum_{i=1}^{13}a_iH_1^{(ai)} + \sum_{i=1}^4H_1^{(bi)} + \sum_{i=1}^{\hspace*{4pt}24\hspace*{4pt}\prime}c_iH_1^{(i)}
\label{H1short}
\end{equation}
where the sum $\sum_{i=1}^{\hspace*{4pt}24\hspace*{4pt}\prime}$ includes $i=8'$ and individual terms of the sums in \eqref{H1short} are defined by
\begin{eqnarray}
 && H_1^{(ai)} = - \frac{v^2}{2}K_{12}^{(i)},\quad H_1^{(b1)} = H_1^{(b2)} = H_1^{(b3)} = 0\\
  \label{H1ij}
 && H_1^{(1)} = H_1^{(2)} = \cdots = H_1^{(8)} = H_1^{({8'})}= 0, \quad H_1^{(15)} = H_1^{(16)} = H_1^{(17)} = 0\nonumber\\
\end{eqnarray}
while the remaining terms $H_1^{(b4)}, H_1^{(9)},\cdots,H_1^{(14)}, H_1^{(18)},\cdots,H_1^{(24)}$ are explicitly given in \eqref{H1}.

The formula \eqref{Hform1} provides a Hamiltonian form of our two-component system \eqref{2comp}
\begin{eqnarray}
&&  u_t = v \nonumber\\
&& v_t = \frac{1}{\Delta}\left(\sum_{i=1}^{13}a_iq^{(ai)} + \sum_{i=1}^4b_iq^{(bi)} + \sum_{i=1}^{\hspace*{4pt}24\hspace*{4pt}\prime}c_iq^{(i)}\right) \equiv \frac{q}{\Delta}
\label{q^i}
\end{eqnarray}
where we have used the short-hand notation
\begin{eqnarray}
&&\hspace*{-9.3pt} q^{(a1)} = v_1^2u_{22} + v_2^2u_{11} - 2v_1v_2u_{12},\quad q^{(a2)} = v_1^2u_{33} + v_3^2u_{11} - 2v_1v_3u_{13} \nonumber\\
&&\hspace*{-9.3pt} q^{(a3)} = v_2^2u_{33} + v_3^2u_{22} - 2v_2v_3u_{23} \nonumber\\
&&\hspace*{-9.3pt} q^{(a4)} = v_1(v_1u_{23} - v_2u_{13}) - v_3(v_1u_{12} - v_2u_{11}) \nonumber\\
&&\hspace*{-9.3pt} q^{(a5)} = v_1(v_2u_{23} - v_3u_{22}) - v_2(v_2u_{13} - v_3u_{12}) \nonumber\\
&&\hspace*{-9.3pt} q^{(a6)} = v_2(v_1u_{33} - v_3u_{13}) - v_3(v_1u_{23} - v_3u_{12}),\quad q^{(a7)} = v_1^2,\quad q^{(a8)} = v_1v_2
\nonumber\\
&&\hspace*{-9.3pt}q^{(a9)} = v_1v_3,\quad  q^{(a10)} = v_2^2,\quad  q^{(a11)} = v_2v_3,\quad q^{(a12)} = v_3^2,\quad q^{(a13)} = 0  \nonumber\\
&&\hspace*{-9.3pt} q^{(b1)} = - \{v_{1}(u_{12}u_{23} - u_{13}u_{22}) - v_{2}(u_{11}u_{23} - u_{12}u_{13}) + v_{3}(u_{11}u_{22} - u_{12}^2)\}\nonumber\\
&&\hspace*{-9.3pt} q^{(b2)} = - \{v_{1}(u_{12}u_{33} - u_{13}u_{23}) - v_{2}(u_{11}u_{33} - u_{13}^2) + v_{3}(u_{11}u_{23} - u_{12}u_{13})\}\nonumber\\
&&\hspace*{-9.3pt} q^{(b3)} = - \{v_{1}(u_{22}u_{33} - u_{23}^2) - v_{2}(u_{12}u_{33} - u_{13}u_{23}) + v_{3}(u_{12}u_{23} - u_{13}u_{22})\}\nonumber\\
&&\hspace*{-20.6pt} q^{(b4)} = - \{u_{11}(u_{22}u_{33} - u_{23}^2) - u_{12}(u_{12}u_{33} - u_{13}u_{23}) + u_{13}(u_{12}u_{23} - u_{13}u_{22})\}\nonumber\\
\label{qai}
\end{eqnarray}

\begin{eqnarray}
&&\hspace*{-9.3pt} q^{(1)} = - (v_{1}u_{12}-v_{2}u_{11}),\quad q^{(2)} = - (v_{1}u_{13}-v_{3}u_{11})  \nonumber\\
&&\hspace*{-9.3pt} q^{(3)} = - (v_{1}u_{22} - v_{2}u_{12}),\quad q^{(4)} = - (v_{1}u_{23}-v_{2}u_{13})
\nonumber\\
&&\hspace*{-9.3pt} q^{(5)} = - (v_{2}u_{23}-v_{3}u_{22}),\quad  q^{(6)} =  - (v_{1}u_{33}-v_{3}u_{13})
\nonumber\\
&&\hspace*{-9.3pt} q^{(7)} = - (v_{2}u_{33}-v_{3}u_{23}),\quad  q^{(8)} = - (v_{2}u_{13}-v_{3}u_{12}) \nonumber\\
&&\hspace*{-9.3pt} q^{({8'})} = - (v_{1}u_{23}-v_{3}u_{12}),\quad q^{(9)} = - (u_{11}u_{23}-u_{12}u_{13}) \nonumber\\
&&\hspace*{-9.3pt} q^{(10)} = - (u_{12}u_{23}-u_{13}u_{22}),\quad q^{(11)} = - (u_{12}u_{33}-u_{13}u_{23})
\nonumber\\
&&\hspace*{-9.3pt} q^{(12)} = - (u_{11}u_{22}-u_{12}^2),\quad q^{(13)} = - (u_{11}u_{33}-u_{13}^2)  \nonumber\\
&&\hspace*{-9.3pt} q^{(14)} = - (u_{22}u_{33}-u_{23}^2),\quad q^{(15)} = - v_{1},\quad  q^{(16)} =  - v_{2},\quad q^{(17)} = - v_{3}
 \nonumber\\
&&\hspace*{-9.3pt} q^{(18)} = - u_{11},\quad q^{(19)} = - u_{12},\quad q^{(20)} = - u_{13},\quad q^{(21)} = - u_{22}\nonumber\\
&&\hspace*{-9.3pt} q^{(22)} = - u_{23},\quad q^{(23)} = - u_{33},\quad q^{(24)} = - 1.
\label{qi}
\end{eqnarray}

An independent check of the Hamiltonian form \eqref{Hform1} for the two component system \eqref{2comp} is conveniently performed with the aid of the relations
\begin{equation}
 \delta_uH_1^{(ai)} - K_{11}v = - q,\quad  \delta_vH_1 = - v\sum_{i=1}^{13}a_iK_{12}^{(i)}\equiv - v K_{12},\quad K_{12} = - \Delta
 \label{rel}
\end{equation}
where $q$ denotes the numerator of the right-hand side of the second equation $v_t=q/\Delta$ in \eqref{q^i}. Here $\delta_u$ and $\delta_v$  are the Euler-Lagrange operators with respect to $u$ and $v$, respectively, \cite{olv} closely related to variational derivatives of the Hamiltonian functional.
The first set of relations \eqref{rel} can be easily checked for the corresponding terms in the definitions \eqref{H1} of $H_1$, \eqref{KijP} and \eqref{KijP3} for $K_{11}$ and \eqref{qai}, \eqref{qi} for $q$ with the result
\begin{eqnarray}
&& \delta_uH_1^{(ai)} - K_{11}^{(ai)}v = - q^{(ai)},\quad \delta_uH_1^{(bi)} - K_{11}^{(bi)}v = - q^{(bi)}\nonumber\\
&& \delta_uH_1^{(i)} - K_{11}^{(i)}v = - q^{(i)}.
 \label{rel_i}
\end{eqnarray}

Using these relations and the definition \eqref{hamilton1} of the Hamiltonian operator $J_0$ in the Hamiltonian system \eqref{Hform1} we obtain
\begin{eqnarray}
&& \left(
\begin{array}{c}
 u_t\\
 v_t
\end{array}
\right) = \left(
    \begin{array}{cc}
   0 & - K_{12}^{-1}\\
   K_{12}^{-1} & K_{12}^{-1}K_{11}K_{12}^{-1}
      \end{array}
    \right)\left(
\begin{array}{c}
 \delta_u H_1\\
 \delta_v H_1
\end{array}
\right) \nonumber\\
&& =
\left(\begin{array}{c}
 - K_{12}^{-1}\delta_v H_1 \\
  K_{12}^{-1}(\delta_u H_1 - K_{11}v)
\end{array}\right) = \left(\begin{array}{c}
 v\\[3pt] \displaystyle
 \left(\frac{q}{\Delta}\right)
\end{array}
\right).
 \label{check}
\end{eqnarray}
which coincides with our original system \eqref{q^i}.

\section{Symmetry condition in a skew-factorized form}
\setcounter{equation}{0}
\label{symmetry}

Symmetry condition is the differential compatibility condition of \eqref{1comp} and the Lie equation $u_\tau = \varphi$, where
$\varphi$ is the symmetry characteristic and $\tau$ is the group parameter. It has the form of Fr\'echet derivative (linearization) of equation \eqref{1comp}.
To have it in a more compact form, we introduce linear differential operators
\begin{eqnarray}
&& L_{ij(k)} = u_{jk}D_i - u_{ik}D_j = - L_{ji(k)}\quad\Longrightarrow L_{ii(k)} = 0,
\label{Lij(k)}\\
&& L_{ij(k)} + L_{ki(j)} + L_{jk(i)} = 0,\quad D_lL_{ij(k)} - D_kL_{ij(l)} =  L_{ij(k)}D_l - L_{ij(l)}D_k\nonumber\\
&& L_{ij(l)}D_k + L_{jk(l)}D_i + L_{ki(l)}D_j =0
\end{eqnarray}
where $i,j,k = 1,2,3,t$. For example,
\begin{eqnarray}
&& L_{12(1)} = u_{12}D_1 - u_{11}D_2,\quad L_{12(2)} = u_{22}D_1 - u_{12}D_2 \nonumber
\\ && L_{12(t)} = u_{2t}D_1 - u_{1t}D_2,\quad L_{12(3)} = u_{23}D_1 - u_{13}D_2 .
\label{Lij(k)exmp}
\end{eqnarray}

In the particular case of equation \eqref{1comp} being a quadratic form in $u_{ij}$, we set $b_{i}=0$, $a_i=0$ for $i=1,\dots 6$ and the symmetry condition with the use of \eqref{Lij(k)} becomes
\begin{eqnarray}
 && \bigl\{a_7(L_{t1(1)}D_t - L_{t1(t)}D_1) +  a_8(L_{t1(2)}D_t - L_{t1(t)}D_2)\nonumber\\
 &&\mbox{} + a_9(L_{t1(3)}D_t - L_{t1(t)}D_3) + a_{10}(L_{t2(2)}D_t - L_{t2(t)}D_2)\nonumber\\
 &&\mbox{} +  a_{11}(L_{t2(3)}D_t - L_{t2(t)}D_3) +  a_{12}(L_{t3(3)}D_t - L_{t3(t)}D_3) \nonumber\\
 &&\mbox{} + c_1(L_{12(1)}D_t - L_{12(t)}D_1) + c_2(L_{13(1)}D_t - L_{13(t)}D_1)\nonumber\\
 &&\mbox{} + c_3(L_{12(2)}D_t - L_{12(t)}D_2) + c_4(L_{12(3)}D_t - L_{12(t)}D_3)\nonumber\\
 &&\mbox{}  + c_5(L_{23(2)}D_t - L_{23(t)}D_2) + c_6(L_{13(3)}D_t - L_{13(t)}D_3)\nonumber\\
 &&\mbox{} + c_7(L_{23(3)}D_t - L_{23(t)}D_3) + c_8(L_{23(1)}D_t - L_{23(t)}D_1)\nonumber\\
 &&\mbox{} + c_{8'}(L_{13(2)}D_t - L_{13(t)}D_2) + c_9(L_{12(3)}D_1 - L_{12(1)}D_3)\nonumber\\
 &&\mbox{} + c_{10}(L_{23(2)}D_1 - L_{23(1)}D_2) + c_{11}(L_{23(3)}D_1 - L_{23(1)}D_3)\nonumber\\
 &&\mbox{} + c_{12}(L_{12(2)}D_1 - L_{12(1)}D_2) + c_{13}(L_{13(3)}D_1 - L_{13(1)}D_3)\nonumber\\
 &&\mbox{} + c_{14}(L_{23(3)}D_2 - L_{23(2)}D_3) + a_{13}D_t^2  + c_{15}D_tD_1 + c_{16}D_tD_2\nonumber\\
 &&\mbox{} + c_{17}D_tD_3 + c_{18}D_1^2 + c_{19}D_1D_2 + c_{20}D_1D_3 + c_{21}D_2^2 + c_{22}D_2D_3 \nonumber\\
 &&\mbox{} + c_{23}D_3^2\bigr\}\varphi = 0.
\label{symcond}
\end{eqnarray}

The linear operator of the symmetry condition for integrable equations of the form \eqref{1comp} should be converted to the "skew-factorized" form
\begin{equation}
 (A_1B_2 - A_2B_1)\varphi = 0
 \label{factorDE}
\end{equation}
where $A_i$ and $B_i$ are first order linear differential operators.
If we introduce two-dimensional vector operators $\vec{R}=(A_1,A_2)$ and $\vec{S}=(B_1,B_2)$, then the skew-factorized form \eqref{factorDE} becomes
the cross (vector) product $(\vec{R}\times\vec{S})\varphi = 0$.
These operators should satisfy the commutator relations
\begin{equation}
  [A_1,A_2] = 0,\quad [A_1,B_2] - [A_2,B_1] = 0,\quad [B_1,B_2] = 0
 \label{commut}
\end{equation}
on solutions of the equation \eqref{1comp}.

It immediately follows that the following two operators also commute on solutions
\begin{equation}
   X_1 = \lambda A_1 + B_1,\quad X_2 = \lambda A_2 + B_2, \qquad [X_1,X_2] = 0
  \label{Lax}
\end{equation}
and therefore constitute Lax representation for equation \eqref{1comp}  with $\lambda$ being a spectral parameter.

Symmetry condition in the form \eqref{factorDE} not only provides Lax pair for equation  \eqref{1comp}
but also leads directly to recursion relations for symmetries
\begin{equation}
  A_1\tilde{\varphi} = B_1\varphi,\quad A_2\tilde{\varphi} = B_2\varphi
 \label{recurs}
\end{equation}
where $\tilde{\varphi}$ is a symmetry if $\varphi$ is also a symmetry and vice versa.
Indeed, equations \eqref{recurs} together with \eqref{commut} imply $(A_1B_2 - A_2B_1)\varphi = [A_1,A_2]\tilde{\varphi}=0$,
so $\varphi$ is a symmetry characteristic. Moreover, due to \eqref{recurs}
\[(A_1B_2 - A_2B_1)\tilde{\varphi} = \bigl([A_1,B_2] - [A_2,B_1] + B_2A_1 - B_1A_2\bigr)\tilde{\varphi} = [B_2,B_1]\varphi = 0\]
which shows that $\tilde{\varphi}$ satisfies the symmetry condition \eqref{factorDE} and hence is also a symmetry. Thus, $\varphi$ is a symmetry whenever so is $\tilde{\varphi}$ and vice versa.
The equations \eqref{recurs} define an auto-B\"acklund transformation between the symmetry conditions written for $\varphi$ and $\tilde{\varphi}$.
Hence, the auto-B\"acklund transformation for the symmetry condition is nothing else than a recursion operator.

We note that the skew-factorized form \eqref{factorDE} and the properties \eqref{commut} of the operators $A_i$ and $B_i$ remain invariant under the simultaneous interchange $A_1\leftrightarrow B_1$ and $A_2\leftrightarrow B_2$.

Our procedure extends A. Sergyeyev's method for constructing recursion operators \cite{Artur}. Namely, unlike \cite{Artur}, we start with the
skew-factorized form of the symmetry condition and extract from there a ``special'' Lax pair instead of building it from a previously known Lax pair.
After that we construct a recursion operator from this newly found Lax pair using a special case of Proposition 1 from \cite{Artur}.

All nonlinear heavenly equations listed in \cite{ferdub}, which describe (anti-)self-dual gravity, can be treated in a unified way according to this approach.

\textit{Second heavenly equation} $u_{tt}u_{11} - u_{t1}^2 + u_{t2} + u_{t3} = 0$ has the symmetry condition of the form
\begin{equation}
  \{L_{t1(1)}D_t - L_{t1(t)}D_1 + D_2D_t + D_3D_1\}\varphi = 0.
  \label{symheav2}
\end{equation}
It has the skew-factorized form \eqref{factorDE} with the operators $A_1 = D_t$, $A_2=D_1$, $B_1 = L_{t1(t)} - D_3$, $B_2 = L_{t1(1)} + D_2$ satisfying
conditions \eqref{commut}. According to \eqref{Lax} the Lax pair has the form $X_1 = \lambda D_t + L_{t1(t)} - D_3$, $X_2 = \lambda D_1 +  L_{t1(1)} + D_2$ and  \eqref{recurs} yields the recursions for symmetries $D_t\tilde{\varphi} = ( L_{t1(t)} - D_3)\varphi$, $D_1\tilde{\varphi} = (L_{t1(1)} + D_2)\varphi$.

\textit{First heavenly equation} in the evolutionary form $(u_{tt} - u_{11})u_{23} - (u_{t3} + u_{13})(u_{t2} - u_{12}) = 1$ has the symmetry condition
\begin{equation}
  \{L_{t2(t)}D_3 - L_{t2(3)}D_t + L_{23(1)}D_t - L_{23(t)}D_1 + L_{12(3)}D_1 - L_{12(1)}D_3\}\varphi = 0
  \label{symheav1}
\end{equation}
with the skew-factorized form composed from the operators $A_1 = D_t - D_1$, $A_2 = - D_3$, $B_1 = L_{t2(t)} - L_{12(1)} - L_{t1(2)}$, $B_2 = L_{t2(3)} + L_{12(3)}$ which satisfy conditions \eqref{commut}. Lax pair \eqref{Lax} reads $X_1 = \lambda (D_t - D_1) + L_{t2(t)} - L_{12(1)} - L_{t1(2)}$,
$X_2 = - \lambda D_3 + L_{t2(3)} + l_{12(3)}$ while the recursion relations \eqref{recurs} become $(D_t - D_1)\tilde{\varphi} = (L_{t2(t)} - L_{12(1)} - L_{t1(2)})\varphi$ and $- D_3\tilde{\varphi} = (L_{t2(3)} + L_{12(3)})\varphi$.

\textit{Modified heavenly equation} $u_{1t}u_{2t} - u_{tt}u_{12} + u_{13} = 0$ has the symmetry condition $(L_{t2(1)}D_t - L_{t2(t)}D_1 - D_1D_3)\varphi = 0$.
Its skew-factorized form is constructed from the operators $A_1 = D_t$, $A_2 = D_1$, $B_1 = L_{t2(t)} + D_3$, $B_2 = L_{t2(1)}$ obviously satisfying conditions \eqref{commut}. The Lax pair \eqref{Lax} is formed by $X_1 = \lambda D_t + L_{t2(t)} + D_3$ and $X_2 = \lambda D_1 + L_{t2(1)}$. Recursions \eqref{recurs}
have the form $D_t\tilde{\varphi} = (L_{t2(t)} + D_3)\varphi$, $D_1\tilde{\varphi} = L_{t2(1)}\varphi$.

\textit{Husain equation} in the evolutionary form $u_{tt} + u_{11} + u_{t2}u_{13} - u_{t3}u_{12} = 0$ has the symmetry condition
$(L_{23(1)}D_t - L_{23(t)}D_1 + D_t^2 + D_1^2)\varphi = 0$. Its skew-factorized form is constituted by the operators $A_1 = D_t$, $A_2 = D_1$,
$B_1 = L_{23(t)} - D_1$, $B_2 = L_{23(1)} + D_t$ satisfying all conditions \eqref{commut}. Lax pair \eqref{Lax} becomes $X_1 = \lambda D_t + L_{23(t)} - D_1$, $X_2 = \lambda D_1 + L_{23(1)} + D_t$ while the recursions \eqref{recurs} read $D_t\tilde{\varphi} = (L_{23(t)} - D_1)\varphi$,
$D_1\tilde{\varphi} = (L_{23(1)} + D_t)\varphi$.

\textit{General heavenly equation} in the evolutionary form
\begin{equation}
(\beta + \gamma)(u_{t2}u_{t3} - u_{tt}u_{23} + u_{11}u_{23} - u_{12}u_{13}) + (\gamma - \beta)(u_{t2}u_{13} - u_{t3}u_{12}) = 0
 \label{genheav}
\end{equation}
has the symmetry condition
\begin{eqnarray}
&& \{(\beta + \gamma)(L_{t3(t)}D_2 - L_{t3(2)}D_t + L_{12(3)}D_1 - L_{12(1)}D_3)\nonumber\\
&&\mbox{} + (\gamma - \beta)(L_{23(1)}D_t - L_{23(t)}D_1)\}\varphi = 0.
 \label{symgen}
\end{eqnarray}
The  skew-factorized form of \eqref{symgen} is achieved with the following operators $A_1 =\displaystyle\frac{1}{u_{23}}L_{t2(3)}$,
$A_2 =\displaystyle\frac{1}{u_{23}}L_{12(3)}$, $B_1 = \displaystyle\frac{1}{u_{23}}\{(\beta-\gamma)L_{t3(2)} + (\beta+\gamma)L_{13(2)}\}$, \\[1mm]$B_2 = \displaystyle \frac{\beta+\gamma}{u_{23}}L_{t3(2)}$. We have checked that these operators satisfy the conditions \eqref{commut}. The Lax pair \eqref{Lax} becomes
$X_1 =  \displaystyle\frac{\lambda}{u_{23}}L_{t2(3)} + \displaystyle\frac{1}{u_{23}}\{(\beta-\gamma)L_{t3(2)} + (\beta+\gamma)L_{13(2)}\}$,
$X_2 = \displaystyle \frac{\lambda}{u_{23}}L_{12(3)}  + \displaystyle \frac{\beta+\gamma}{u_{23}}L_{t3(2)}$. Recursion relations \eqref{recurs} have the form
\begin{eqnarray}
&& \displaystyle\frac{1}{u_{23}}L_{t2(3)}\tilde{\varphi} =  \displaystyle\frac{1}{u_{23}}\{(\beta-\gamma)L_{t3(2)} + (\beta+\gamma)L_{13(2)}\}\varphi\nonumber\\ && \displaystyle \frac{1}{u_{23}}L_{12(3)}\tilde{\varphi} = \displaystyle \frac{\beta+\gamma}{u_{23}}L_{t3(2)}\varphi.
 \label{recursgenheav}
\end{eqnarray}

\section{Symmetry condition, integrability and recursion}
\setcounter{equation}{0}
\label{sec-integr}

A regular way for arriving at skew-factorized forms of the symmetry condition \eqref{symcond}, for the particular case of equation \eqref{1comp} being a quadratic form of $u_{ij}$, is based on the following relations between the operators $L_{ij(k)}$
\begin{eqnarray}
   &&\hspace*{-18pt} L_{ij(k)}D_l - L_{ij(l)}D_k = L_{ij(k)}\frac{1}{u_{jk}}L_{lk(j)} + D_j\frac{1}{u_{jk}}(u_{jk}u_{il} - u_{ik}u_{jl})D_k
 \label{rel1}
\\ &&\hspace*{-18pt} L_{ij(k)}D_l - L_{ij(l)}D_k = L_{lk(j)}\frac{1}{u_{jk}}L_{ij(k)} + D_k\frac{1}{u_{jk}}(u_{jk}u_{il} - u_{ik}u_{jl})D_j
 \label{rel2}
\\ &&\hspace*{-18pt} L_{ij(k)}D_l - L_{ij(l)}D_k = L_{ij(l)}\frac{1}{u_{jl}}L_{lk(j)} + D_j\frac{1}{u_{jl}}(u_{jk}u_{il} - u_{ik}u_{jl})D_l
 \label{rel3}
\\ &&\hspace*{-18pt} L_{ij(k)}D_l - L_{ij(l)}D_k = L_{li(j)}\frac{1}{u_{ij}}L_{kj(i)} - L_{ki(j)}\frac{1}{u_{ij}}L_{lj(i)}\nonumber
\\ &&\hspace*{-18pt}\mbox{} + D_i \frac{1}{u_{ij}} (u_{jk}u_{il} - u_{ik}u_{jl})D_j.
 \label{rel4}
\end{eqnarray}
We note that the expression in parentheses in the last term of all these four relations is precisely the group of terms in the equation \eqref{1comp}
which corresponds to the terms $(L_{ij(k)}D_l - L_{ij(l)}D_k)\varphi$ in the symmetry condition \eqref{symcond}, so that the last terms in all these relations vanish on solutions of \eqref{1comp}.

Keeping different groups of terms in \eqref{1comp}, we obtain skew-factorized forms of the symmetry condition \eqref{symcond} determined by
the operators $A_i, B_i$  listed below which satisfy all the conditions \eqref{commut}, as shown at the end of this section. Using \eqref{Lax} and \eqref{recurs} we immediately obtain the Lax pair and recursion relations, respectively, for all these equations.

For the equation
\begin{eqnarray}
&& a_{11}(u_{tt}u_{23}-u_{t2}u_{t3}) + c_4(u_{t1}u_{23}-u_{t2}u_{13}) + c_5 (u_{t2}u_{23} - u_{t3}u_{22})
 \label{eqi}\\
&&\mbox{}  + c_8 (u_{t2}u_{13} - u_{t3}u_{12}) + c_9(u_{11}u_{23} - u_{12}u_{13}) + c_{10}(u_{12}u_{23}-u_{13}u_{22}) = 0\nonumber
\end{eqnarray}
\begin{eqnarray}
&& A_1 = \frac{1}{u_{23}}L_{t2(3)},\quad  B_1 = \frac{1}{u_{23}}\{(c_4-c_8)L_{t3(2)} + c_9L_{13(2)} + c_{10}L_{23(2)}\}\nonumber\\
&& A_2 = - \frac{1}{u_{23}}L_{12(3)},\quad B_2 = \frac{1}{u_{23}}(c_5L_{23(2)} + c_8L_{13(2)} + a_{11}L_{t3(2)}).
\label{ABi}
\end{eqnarray}

For the equation
\begin{eqnarray}
&& a_{11}(u_{tt}u_{23}-u_{t2}u_{t3}) + c_4(u_{t1}u_{23}-u_{t2}u_{13}) + c_7(u_{t2}u_{33}-u_{t3}u_{23})
 \label{eqii}\\
&&\mbox{}  + c_8 (u_{t2}u_{13} - u_{t3}u_{12}) + c_9(u_{11}u_{23} - u_{12}u_{13}) + c_{11}(u_{12}u_{33}-u_{13}u_{23}) = 0\nonumber
\end{eqnarray}
\begin{eqnarray}
&& A_1 = \frac{1}{u_{23}}L_{t3(2)},\quad B_1 = \frac{1}{u_{23}}(c_8L_{t2(3)} + c_9L_{12(3)} + c_{11}L_{23(3)})
\label{ABii}\\
&& A_2 = - \frac{1}{u_{23}}L_{13(2)},\quad B_2 =  \frac{1}{u_{23}}\{(c_4-c_8)L_{12(3)} + c_7L_{23(3)} + a_{11}L_{t2(3)}\}.\nonumber
\end{eqnarray}

For the equation
\begin{eqnarray}
 && a_8(u_{tt}u_{12}-u_{t1}u_{t2}) + a_{10}(u_{tt}u_{22}-u_{t2}^2) + a_{11}(u_{tt}u_{23}-u_{t2}u_{t3}) \nonumber\\
 &&\mbox{} + c_7(u_{t2}u_{33}-u_{t3}u_{23}) + c_8(u_{t2}u_{13} - u_{t3}u_{12}) = 0
\label{eqvii}
\end{eqnarray}
\begin{eqnarray}
  && A_1 = \frac{1}{u_{t2}}L_{23(t)},\quad B_1 = \frac{1}{u_{t2}}(a_8L_{t1(2)} + a_{10}L_{t2(2)} + a_{11}L_{t3(2)}) \nonumber\\
  &&  A_2 = - \frac{1}{u_{t2}}L_{t2(t)},\quad B_2 = \frac{1}{u_{t2}}(c_7L_{t3(2)} + c_8L_{t1(2)}).
   \label{ABvii}
\end{eqnarray}

For the equation
\begin{eqnarray}
&& a_{12}(u_{tt}u_{33}-u_{t3}^2) + c_5 (u_{t2}u_{23} - u_{t3}u_{22}) + c_6(u_{t1}u_{33}-u_{t3}u_{13})\nonumber\\
&&\mbox{} + c_7(u_{t2}u_{33}-u_{t3}u_{23}) + c_8(u_{t2}u_{13} - u_{t3}u_{12}) = 0
\label{eqviii}
\end{eqnarray}
\begin{eqnarray}
  &&\hspace*{-3.5pt} A_1 = \frac{1}{u_{t3}}L_{t3(3)},\quad B_1 = - \frac{1}{u_{t3}}L_{23(t)}
     \label{ABviii} \\
  &&\hspace*{-3.5pt} A_2 = \frac{1}{u_{t3}}(c_5L_{t2(3)} + c_8L_{t1(3)}),\quad B_2 = \frac{1}{u_{t3}}(a_{12}L_{t3(t)} + c_6L_{13(t)} + c_7L_{23(t)})\nonumber
\end{eqnarray}

For the equation
\begin{eqnarray}
&& a_7(u_{tt}u_{11}-u_{t1}^2) + a_8(u_{tt}u_{12}-u_{t1}u_{t2}) + a_9(u_{tt}u_{13} - u_{t1}u_{t3})\nonumber\\
&&\mbox{} + c_1(u_{t1}u_{12}-u_{t2}u_{11}) + c_3(u_{t1}u_{22} - u_{t2}u_{12}) + c_4(u_{t1}u_{23}-u_{t2}u_{13}) = 0\nonumber\\
\label{eqx}
\end{eqnarray}
\begin{eqnarray}
  && A_1 = \frac{1}{u_{t1}}L_{t1(t)} ,\quad B_1 = \frac{1}{u_{t1}}(c_1L_{t1(1)} + c_3L_{t2(1)} + c_4L_{t3(1)})\nonumber\\
  && A_2 = - \frac{1}{u_{t1}}L_{12(t)} ,\quad B_2 = \frac{1}{u_{t1}}(a_7L_{t1(1)} + a_8L_{t2(1)} + a_9L_{t3(1)}).
   \label{ABx}
\end{eqnarray}

We can obtain skew-factorized forms of symmetry conditions for many more equations of the type \eqref{1comp} by using
permutations of indices $1,2,3,t$ with an appropriate permutation of coefficients which leave the equation \eqref{1comp} invariant.
Such permutations will however do change the skew factorized forms of the symmetry conditions.
For example, the transposition of indices $1\leftrightarrow 2$ will leave the equation invariant if it is accompanied with the following transpositions of coefficients
\begin{eqnarray}
&& a_7\leftrightarrow a_{10},\quad a_9\leftrightarrow a_{11},\quad c_1\leftrightarrow -c_3,\quad c_2\leftrightarrow c_5,\quad c_6\leftrightarrow c_7,
\quad  c_4\leftrightarrow -c_4 \nonumber\\
&& c_8\leftrightarrow c_{8'},\quad c_9\leftrightarrow -c_{10},\quad c_{13}\leftrightarrow c_{14}
\label{permut}
\end{eqnarray}
with all other coefficients unchanged. Applying this transformation to the skew-factorized form for the equation \eqref{eqi}, we obtain a new one
\begin{eqnarray}
&& \left\langle L_{t1(3)}\frac{1}{u_{13}}(c_2L_{13(1)} + c_{8'}L_{23(1)} + a_9L_{t3(1)})\right.\nonumber\\
&&\left. \mbox{} + L_{12(3)}\frac{1}{u_{13}}\{(c_4+c_{8'})L_{t3(1)} + c_9L_{13(1)} + c_{10}L_{23(1)}\}\right\rangle\varphi = 0
 \label{ia}
\end{eqnarray}
which provides the Lax pair and recursions for the equation
\begin{eqnarray}
&&\hspace*{-5.5pt} a_9(u_{tt}u_{13} - u_{t1}u_{t3}) + c_2(u_{t1}u_{13} - u_{t3}u_{11}) + c_4(u_{t1}u_{23}-u_{t2}u_{13})
 \label{eqia}\\
&&\hspace*{-5.5pt}\mbox{} + c_{8'} (u_{t1}u_{23} - u_{t3}u_{12}) + c_9(u_{11}u_{23} - u_{12}u_{13}) + c_{10}(u_{12}u_{23}-u_{13}u_{22}) = 0.\nonumber
\end{eqnarray}

Some of the equations listed above are not independent in the sense that they are related by some permutation of indices. For example, our second equation here \eqref{eqii} and the corresponding operators $A_i, B_i$ in \eqref{ABii}, which determine the Lax pair and recursion relations, can be obtained from the first equation \eqref{eqi} and its operators \eqref{ABi} by the transposition of indices $2\leftrightarrow 3$ and the permutation of the coefficients $c_5\leftrightarrow -c_7$, $c_8\leftrightarrow (c_4-c_8)$ and $c_{10}\leftrightarrow -c_{11}$ with all other coefficients (including $c_4$) unchanged.

To see that  conditions \eqref{commut} are satisfied for any operators arising from the skew-factorized form of the symmetry condition \eqref{symcond}, we note that this form should follow from a linear combination of such pairs of terms in the symmetry condition \eqref{symcond}
\begin{equation}
p(L_{ij(k)}D_l - L_{ij(l)}D_k) + q(L_{mj(k)}D_n - L_{mj(n)}D_k),
\label{L_pq}
\end{equation}
with constant $p,q$, which are simultaneously factorized on solutions of the corresponding equations according the formula \eqref{rel1}
\begin{eqnarray}
   &&\hspace*{-18pt} L_{ij(k)}D_l - L_{ij(l)}D_k = L_{ij(k)}\frac{1}{u_{jk}}L_{lk(j)} + D_j\frac{1}{u_{jk}}(u_{jk}u_{il} - u_{ik}u_{jl})D_k\nonumber\\
   &&\hspace*{-18pt} L_{mj(k)}D_n - L_{mj(n)}D_k = L_{mj(k)}\frac{1}{u_{jk}}L_{nk(j)} + D_j\frac{1}{u_{jk}}(u_{jk}u_{mn} - u_{mk}u_{jn})D_k.\nonumber\\
   \label{fact}
\end{eqnarray}
Here the factors $D_j(1/u_{jk})(E_{p,q})D_k$ are the same in both formulas with the exception of factors $E_p$, $E_q$, where
\begin{eqnarray}
&& E_p = u_{jk}u_{il} - u_{ik}u_{jl}, \quad E_q = u_{jk}u_{mn} - u_{mk}u_{jn}
   \label{E_pq}
\end{eqnarray}
constitute the parts of the equation $E_{pq} = pE_p + qE_q = 0$ which implies the symmetry condition \eqref{L_pq}. Then on solutions of the equation $E_{pq} = 0$ we have
\begin{eqnarray}
&& p(L_{ij(k)}D_l - L_{ij(l)}D_k) + q(L_{mj(k)}D_n - L_{mj(n)}D_k) \nonumber\\
&& = pL_{ij(k)}\frac{1}{u_{jk}}L_{lk(j)} + qL_{mj(k)}\frac{1}{u_{jk}}L_{nk(j)}
 \label{factor}
\end{eqnarray}
which yields skew-factorized form with the operators
\begin{eqnarray*}
&&\hspace*{-11pt} A_1 = \frac{1}{u_{jk}}L_{ij(k)},\quad B_2 = \frac{1}{u_{jk}}L_{lk(j)},\quad A_2 = - \frac{1}{u_{jk}}L_{mj(k)},\quad %\nonumber\\&&
B_1 = \frac{1}{u_{jk}}L_{nk(j)}.
% \label{A_iB_i}
\end{eqnarray*}
It is easy to check that for these operators we have $[A_1,A_2]=0$ and $[B_1,B_2]=0$ identically satisfied, which justifies the factors $1/u_{jk}$ in $A_i$, while $[A_1,B_2] - [A_2,B_1] = 0$ holds on solutions of the equation $E_{pq} = 0$ due to the identity
\begin{equation*}
[A_1,B_2] - [A_2,B_1] = \frac{1}{u_{jk}}\left\{D_k\left(\frac{E_{pq}}{u_{jk}}\right)D_j - D_j\left(\frac{E_{pq}}{u_{jk}}\right)D_k\right\}
\end{equation*}
where in the braces we have the products of operators.
More general skew-factorized forms of the symmetry condition arise as suitable linear combinations of the equations of the form \eqref{factor}.

\section{Recursion operators in $2\times 2$ matrix form}
\setcounter{equation}{0}
\label{recursoper}

To construct new two-component bi-Hamiltonian systems we need recursion operators in a $2\times 2$ matrix form. We demonstrate the procedure in detail for our first equation \eqref{eqi} admitting skew-factorized symmetry condition with recursion relations \eqref{recurs} determined by the operators \eqref{ABi}
\begin{eqnarray}
&& u_{23}\tilde{\varphi}_t - u_{t3}\tilde{\varphi}_2 = (c_4-c_8)(u_{23}\varphi_t - u_{t2}\varphi_3) + (c_9L_{13(2)} + c_{10}L_{23(2)})\varphi\nonumber\\
&& - L_{12(3)}\tilde{\varphi} = (c_5L_{23(2)} + c_8L_{13(2)})\varphi + a_{11}(u_{23}\varphi_t - u_{t2}\varphi_3).
 \label{recurs1}
\end{eqnarray}
Lax pair for the equation \eqref{eqi} reads
\begin{eqnarray}
&& X_1 = \frac{\lambda}{u_{23}}L_{t2(3)} + \frac{1}{u_{23}}\{(c_4-c_8)L_{t3(2)} + c_9L_{13(2)} + c_{10}L_{23(2)}\}\nonumber\\
&& X_2 = - \frac{\lambda}{u_{23}}L_{12(3)} + \frac{1}{u_{23}}(c_5L_{23(2)} + c_8L_{13(2)} + a_{11}L_{t3(2)}).
\label{Lax1}
\end{eqnarray}

In a two-component form the equation \eqref{eqi} becomes
\begin{eqnarray}
&&\hspace*{-3.3pt} u_t = v,
  \label{eqi2comp}\\
&&\hspace*{-3.3pt}  v_t = \frac{q}{\Delta} = \frac{1}{a_{11}u_{23}}\left(a_{11}q^{(a11)} + c_4q^{(4)} + c_5q^{(5)} + c_8q^{(8)} + c_9q^{(9)} + c_{10}q^{(10)} \right)\nonumber
\end{eqnarray}
where according to \eqref{qai}, \eqref{qi} we have
\begin{eqnarray}
&& q^{(a11)} = v_2v_3,\quad q^{(4)} = -(v_1u_{23} - v_2u_{13}),\quad q^{(5)} = -(v_2u_{23} - v_3u_{22}),\nonumber\\
&& q^{(8)} = -(v_2u_{13} - v_3u_{12}),\quad q^{(9)} = -(u_{11}u_{23} - u_{12}u_{13})\nonumber\\
&& q^{(10)} = -(u_{12}u_{23} - u_{13}u_{22}).
\label{q}
\end{eqnarray}

Lie equations in a two-component form become $u_\tau = \varphi$, $v_\tau = \psi$, so that $u_t = v$ implies $\varphi_t = \psi$.
We define two-component symmetry characteristic $(\varphi,\psi)^T$ (where $^T$ means transposed matrix) with $\psi=\varphi_t$ and $(\tilde{\varphi},\tilde{\psi})^T$ with $\tilde{\psi}=\tilde{\varphi}_t$ for the original and transformed symmetries, respectively.
In this form recursions \eqref{recurs1} become
\begin{eqnarray}
&& u_{23}\tilde{\psi} - v_3\tilde{\varphi}_2 = (c_4-c_8)(u_{23}\psi - v_2\varphi_3) + (c_9L_{13(2)} + c_{10}L_{23(2)})\varphi\nonumber\\
\label{1stcomp}\\
&& - L_{12(3)}\tilde{\varphi} = (c_5L_{23(2)} + c_8L_{13(2)})\varphi + a_{11}(u_{23}\psi - v_2\varphi_3).
 \label{2ndcomp}
\end{eqnarray}
We first solve \eqref{2ndcomp} with respect to $\tilde{\varphi}$
\begin{eqnarray}
 \tilde{\varphi} = - L_{12(3)}^{-1}(c_5L_{23(2)} + c_8L_{13(2)} - a_{11}v_2D_3)\varphi - a_{11}L_{12(3)}^{-1}u_{23}\psi
 \label{1stsolve}
\end{eqnarray}
then solve \eqref{1stcomp} with respect to $\tilde{\psi}$
\begin{eqnarray*}
&&\hspace*{-10pt} \tilde{\psi} = \frac{1}{u_{23}}\{v_3D_2  \tilde{\varphi} + (c_8 - c_4)v_2D_3\varphi + (c_9L_{13(2)} + c_{10}L_{23(2)})\varphi\} + (c_4-c_8)\psi
\end{eqnarray*}
and use here $\tilde{\varphi}$ from \eqref{1stsolve}. By definition, we require the operator inverse to $L_{12(3)}$ to satisfy the relation
$L_{12(3)}^{-1}L_{12(3)} = 1$.

We present the result in the matrix form using a $2\times 2$ matrix recursion operator $R$
\begin{equation}
 \left(\begin{array}{c}
  \tilde{\varphi}\\
  \tilde{\psi}
 \end{array}\right) = R\left(\begin{array}{c}
  \varphi\\
  \psi
 \end{array}\right), \quad R = \left(\begin{array}{cc}
  R_{11} & - a_{11}L_{12(3)}^{-1}u_{23}\\
  R_{21} & - a_{11}\displaystyle\frac{v_3}{u_{23}}D_2L_{12(3)}^{-1}u_{23} + c_4 - c_8
 \end{array}\right)
 \label{R}
\end{equation}
with the matrix elements
\begin{eqnarray*}
 && R_{11} = - L_{12(3)}^{-1}(c_5L_{23(2)} + c_8L_{13(2)} - a_{11}v_2D_3)\\
 && R_{21} = \frac{1}{u_{23}}\{(c_8 - c_4)v_2D_3 + c_9L_{13(2)} + c_{10}L_{23(2)}\}\\
 &&\mbox{} -\frac{v_3}{u_{23}}D_2L_{12(3)}^{-1}(c_5L_{23(2)} + c_8L_{13(2)} - a_{11}v_2D_3).
\end{eqnarray*}

We note that the operator $L_{12(3)}^{-1}$ can make sense merely as a \textit{formal} inverse of $L_{12(3)}$. Thus, the recursion relations above are formal as well. The proper interpretation of the quantities like $L_{12(3)}^{-1}$ requires the language of differential coverings (see the recent survey \cite{Kras} and references therein).

\section{Second Hamiltonian representation}
\setcounter{equation}{0}
\label{sec-biHam}

Composing the recursion operator \eqref{R} with the Hamiltonian operator $J_0$ defined in \eqref{hamilton1} we will obtain the second Hamiltonian operator
$J_1 = RJ_0$. For the equation \eqref{eqi2comp}   according to formulas \eqref{Kij} and \eqref{KijP3} we have $K_{12} = - a_{11}u_{23}$,
$K_{11} = a_{11}(v_3D_2 + D_3v_2) - c_4L_{12(3)} - c_5L_{23(2)} - c_8L_{23(1)}$. Expression \eqref{hamilton1} for $J_0$ becomes
\begin{equation}
J_0 = \frac{1}{a_{11}u_{23}}\left(\begin{array}{cc}
    0  & 1 \\
    -1 &\displaystyle \frac{1}{a_{11}}K_{11}\frac{1}{u_{23}}
\end{array}\right).
  \label{J_0}
\end{equation}
The corresponding Hamiltonian density according to \eqref{H1}, \eqref{H1short} reads
\begin{eqnarray}
&& H_1 = a_{11}H_1^{(a11)} + c_9H_1^{(9)} + c_{10}H_1^{(10)}\nonumber\\
&& = a_{11}\frac{v^2}{2}u_{23} + \frac{u}{3}\{c_9(u_{11}u_{23} - u_{12}u_{13}) + c_{10}(u_{12}u_{23} - u_{13}u_{22})\}.
 \label{H1I}
\end{eqnarray}
The equation \eqref{eqi} taken in the two-component form \eqref{eqi2comp}, \eqref{q} can be written now as the Hamiltonian system
\begin{equation}
\left(
\begin{array}{c}
 u_t\\
 v_t
\end{array}
\right) = J_0
\left(
\begin{array}{c}
 \delta_u H_1\\
 \delta_v H_1
\end{array}
\right).
  \label{Hamsys}
\end{equation}

For bi-Hamiltonian system we need a second Hamiltonian operator and corresponding Hamiltonian density.
Performing matrix multiplication $RJ_0$ of the expressions \eqref{R} and \eqref{J_0} we obtain the second Hamiltonian operator
\begin{equation}
 J_1 = \left(\begin{array}{cc}
  L_{12(3)}^{-1}             & - \left(L_{12(3)}^{-1}D_2v_3 + \displaystyle\frac{c_8-c_4}{a_{11}}\right)\displaystyle\frac{1}{u_{23}} \\
  \displaystyle\frac{1}{u_{23}}\left(v_3D_2L_{12(3)}^{-1} + \displaystyle\frac{c_8-c_4}{a_{11}}\right) & J_1^{22}
 \end{array}\right)
 \label{J_1}
\end{equation}
where the entry $J_1^{22}$ is defined by
\begin{eqnarray}
 && J_1^{22} = \frac{1}{a_{11}u_{23}}(c_9L_{13(2)} + c_{10}L_{23(2)})\frac{1}{u_{23}} - \frac{v_3}{u_{23}}D_2L_{12(3)}^{-1}D_2\frac{v_3}{u_{23}}
 \label{J_1^22}
\\
&&\mbox{} + \frac{c_4-c_8}{a_{11} u_{23}}\left\{D_2v_3 + v_3D_2
 - \frac{1}{a_{11}}(c_4L_{12(3)} + c_5L_{23(2)} + c_8L_{23(1)})\right\}\frac{1}{u_{23}}.\nonumber
\end{eqnarray}
The formulas \eqref{J_1} and \eqref{J_1^22} show that operator $J_1$ is manifestly skew symmetric.
A check of the Jacobi identities and compatibility of the two Hamiltonian structures $J_0$ and $J_1$ is straightforward but too lengthy to be presented here. The method of the functional multi-vectors for checking the Jacobi identity and the compatibility of the Hamiltonian operators is developed by P. Olver in \cite{olv}, chapter 7 and has been recently applied for checking bi-Hamiltonian structure of the general heavenly equation \cite{sym} and the first heavenly equation of Pleba\'nski \cite{sy} under the well-founded conjecture that this method is applicable for nonlocal Hamiltonian operators as well.

The next problem is to derive the Hamiltonian density $H_0$ corresponding to the second Hamiltonian operator $J_1$ such that
implies the bi-Hamiltonian representation of the system \eqref{eqi2comp}
\begin{equation}
\left(
\begin{array}{c}
 u_t\\
 v_t
\end{array}
\right) = J_0
\left(
\begin{array}{c}
 \delta_u H_1\\
 \delta_v H_1
\end{array}
\right) = J_1
\left(
\begin{array}{c}
 \delta_u H_0\\
 \delta_v H_0
\end{array}
\right) = \left(
\begin{array}{c}
 v\\ \displaystyle
 \frac{q}{\Delta}
\end{array}
\right)
  \label{be_Ham}
  \end{equation}
where $q/\Delta$ is the right-hand side of the second equation in \eqref{eqi2comp}. Then we may conclude that our system is integrable in the sense of Magri \cite{magri}.

We further assume quadratic dependence of the Hamiltonian $H_0$ on $v$
\begin{equation}
H_0 = a[u]v^2 + b[u]v + c[u]
 \label{assump}
\end{equation}
with the coefficients depending only on $u$ and its partial derivatives. Hence $\delta_vH_0 = 2a[u]v + b[u]$.

\begin{Proposition}
 Bi-Hamiltonian representation \eqref{be_Ham} of the system \eqref{eqi2comp} with the assumption \eqref{assump} is valid under the constraint
 \begin{equation}
 c_8c_{10} = c_5c_9
  \label{constrH0}
 \end{equation}
  with the following Hamiltonian density
\begin{eqnarray}
H_0 = -\frac{\{a_{11}c_8v^2 + (a_{11}c_9u_1 + b_0)v - c_9(c_8-c_4)u_1^2\}u_{23}}{2\{a_{11}c_9 + c_8(c_8-c_4)\}}
\label{H0}
\end{eqnarray}
\end{Proposition}
\underline{\textit{Proof}}.\\[3pt]
Multiplying the first row of $J_1$ by the column of variational derivatives of $H_0$ in \eqref{be_Ham} and applying $L_{12(3)}$ we obtain
\begin{eqnarray}
 \delta_uH_0 = D_2\frac{v_3}{u_{23}}(2a[u]v + b[u]) + L_{12(3)}\left\{\frac{(c_8-c_4)}{a_{11}u_{23}}(2a[u]v + b[u]) + v\right\}
 \label{1stline}
\end{eqnarray}
The second row of the last equation in \eqref{be_Ham} with the use of \eqref{q}, \eqref{J_1}, \eqref{J_1^22} and \eqref{1stline} reads
\begin{eqnarray}
&&\hspace*{-18pt} c_9\left\{2a(u_{23}v_1 - u_{12}v_3) + \left(2vL_{13(2)}\left[\frac{a}{u_{23}}\right] + L_{13(2)}\left[\frac{b}{u_{23}}\right]\right)u_{23}\right\}\nonumber\\
&&\hspace*{-18pt}\mbox{} + c_{10}\left\{2a(u_{23}v_2 - u_{22}v_3) + \left(2vL_{23(2)}\left[\frac{a}{u_{23}}\right] + L_{23(2)}\left[\frac{b}{u_{23}}\right]\right)u_{23}\right\}
\nonumber\\
&&\hspace*{-18pt} \mbox{} - \frac{(c_4-c_8)c_5}{a_{11}}\left\{2a(u_{23}v_2 - u_{22}v_3) + \left(2vL_{23(2)}\left[\frac{a}{u_{23}}\right] + L_{23(2)}\left[\frac{b}{u_{23}}\right]\right)u_{23}\right\}\nonumber\\
&&\hspace*{-18pt} \mbox{} - \frac{(c_4-c_8)c_8}{a_{11}}\left\{2a(u_{23}v_1 - u_{12}v_3) + \left(2vL_{13(2)}\left[\frac{a}{u_{23}}\right] + L_{13(2)}\left[\frac{b}{u_{23}}\right]\right)u_{23}\right\}\nonumber\\
&&\hspace*{-18pt} = - \{c_5(u_{23}v_2 - u_{22}v_3) + c_8(u_{23}v_1 - u_{12}v_3) + c_9L_{12(3)}u_1 + c_{10}L_{23(2)}u_1\}u_{23}.\nonumber\\
\label{2ndline}
 \end{eqnarray}
Here all the dependence on $v$ and its derivatives is explicit, so that we may split \eqref{2ndline} into separate equations containing terms with $v_1$, $v_2$,
$v_3$, $v$ and without $v$  The group of terms with $v_1$ yields
\begin{equation}
 a[u] = - \frac{a_{11}c_8u_{23}}{2\{a_{11}c_9 + c_8(c_8-c_4)\}}.
 \label{a}
\end{equation}
The group of terms with $v_2$ results in the constraint \eqref{constrH0}. With these results the group of terms with $v_3$ vanishes. Since, due to \eqref{a}, $a/u_{23}$ is constant and hence $L_{ij(k)}[a/u_{23}] = 0$, all terms with $v$ vanish, so that the remaining terms in \eqref{2ndline} read
\begin{eqnarray}
 && c_9L_{12(3)}[B+u_1] + c_{10}L_{23(2)}[B+u_1]\nonumber\\
 && = \frac{(c_4-c_8)}{a_{11}}(c_8L_{13(2)}[B] + c_5L_{23(2)}[B]) - c_9L_{23(1)}[B]
 \label{no v}
\end{eqnarray}
where we have defined $B[u] = b[u]/u_{23}$. Applying to \eqref{no v} the constraint \eqref{constrH0} in the form $c_{10}=c_5c_9/c_8$ and the relation
$L_{13(2)} = L_{12(3)} + L_{23(1)}$, we rewrite \eqref{no v} in the form
\begin{eqnarray}
 && (c_8L_{12(3)} + c_5L_{23(2)})\left[c_9(B+u_1) - \frac{c_8(c_4-c_8)}{a_{11}}B\right]\nonumber\\
 && = c_8\left\{\frac{c_8(c_4-c_8)}{a_{11}} - c_9\right\}L_{23(1)}[B].
  \label{eqB}
\end{eqnarray}
An obvious solution to \eqref{eqB} is such $B$ for which the expression in square brackets is a constant, so that the left-hand side of \eqref{eqB} vanishes.
Then $B$ is a linear function of $u_1$  with constant coefficients which also annihilates the right-hand side because $L_{23(1)}[u_1]\equiv 0$. Thus, the solution is $B = \displaystyle \frac{c_9a_{11}u_1 + b_0}{c_8(c_4-c_8) - c_9a_{11}}$, where $b_0$ is an arbitrary constant, which yields
\begin{equation}
b[u] = \frac{(c_9a_{11}u_1 + b_0)u_{23}}{c_8(c_4-c_8) - c_9a_{11}}.
 \label{b}
\end{equation}

Next we come back to equation \eqref{1stline} utilizing our results \eqref{a} and \eqref{b} for $a$ and $b$, respectively, and evaluating
$\delta_u(a[u]v^2 + b[u]v + c[u])$ on the left side. We end up with the result\\[2pt]
$\delta_uc[u] = \displaystyle\frac{(c_8-c_4)c_9}{a_{11}c_9 + c_8(c_8-c_4)}(u_{12}u_{13} - u_{11}u_{23})$ which obviously implies
\begin{equation}
c[u] = \frac{c_9(c_8 - c_4)}{2\{a_{11}c_9 + c_8(c_8-c_4)\}}u_1^2u_{23}.
 \label{c}
\end{equation}
Applying our results \eqref{a}, \eqref{b} and \eqref{c} to our ansatz \eqref{assump} for $H_0$ we obtain the required formula \eqref{H0}.
\\$\Box$

Thus, we have shown that equation \eqref{eqi} in the  two-component form \eqref{eqi2comp} under the constraint \eqref{constrH0} admits bi-Hamiltonian representation \eqref{be_Ham} with the second Hamiltonian operator $J_1$ defined in \eqref{J_1}, \eqref{J_1^22} and the corresponding Hamiltonian density $H_0$ given in \eqref{H0}. In section \ref{sec-biHam2}, in a quite similar way we construct bi-Hamiltonian systems corresponding to other four equations admitting skew-factorized form of the symmetry condition which are listed in section \ref{sec-integr}.

\section{Further new bi-Hamiltonian systems}
\setcounter{equation}{0}
\label{sec-biHam2}

We present here without proofs similar results for recursion operators, first and second Hamiltonian operators and the corresponding Hamiltonian densities for other equations from section \ref{sec-integr} possessing recursions. For all these equations we use the same ansatz $H_0 = a[u]v^2 + b[u]v + c[u]$ for the second Hamiltonian density. The proofs can be performed along the same lines as for the first equation \eqref{eqi} from section \ref{sec-integr}.

\subsection{Equation \eqref{eqii}}

As was mentioned above the results for equation \eqref{eqii} can be obtained from those for equation \eqref{eqi} by interchanging the indices 2 and 3 together with the simultaneous interchange of the coefficients $c_5\leftrightarrow -c_7$, $c_8\leftrightarrow (c_4-c_8)$ and $c_{10}\leftrightarrow -c_{11}$ with all other coefficients (including $c_4$) unchanged.

The Lax pair for equation \eqref{eqii} reads
\begin{eqnarray}
&& X_1 = \frac{\lambda}{u_{23}}L_{t3(2)} + \frac{1}{u_{23}}\{c_8L_{t2(3)} + c_9L_{12(3)} + c_{11}L_{23(3)}\}\nonumber\\
&& X_2 = - \frac{\lambda}{u_{23}}L_{13(2)} + \frac{1}{u_{23}}\{(c_4 - c_8)L_{12(3)} + c_7L_{23(3)} + a_{11}L_{t2(3)}\}.
\label{Laxii}
\end{eqnarray}

The equation \eqref{eqii} in the two-component form becomes
\begin{eqnarray}
&& u_t = v
 \label{eqii2}\\
&& v_t =\frac{q}{\Delta} =\frac{1}{a_{11}u_{23}}\{a_{11}v_2v_3 + c_4(v_2u_{13} - v_1u_{23}) - c_7(v_2u_{33} - v_3u_{23})\nonumber\\
&&\hspace*{-11.3pt}\mbox{} - c_8(v_2u_{13} - v_3u_{12}) + c_9(u_{12}u_{13} - u_{11}u_{23}) - c_{11}(u_{12}u_{33} - u_{13}u_{23})\}.\nonumber
\end{eqnarray}
The recursion operator is obtained from \eqref{R} by the same combined permutation
\begin{equation}
R = \left(\begin{array}{cc}
  R_{11} & - a_{11}L_{13(2)}^{-1}u_{23}\\
  R_{21} & - a_{11}\displaystyle\frac{v_2}{u_{23}}D_3L_{13(2)}^{-1}u_{23} + c_8
 \end{array}\right)
 \label{Req2}
\end{equation}
with the matrix elements
\begin{eqnarray*}
 && R_{11} = - L_{13(2)}^{-1}(c_7L_{23(3)} + (c_4 - c_8)L_{12(3)} - a_{11}v_3D_2)\\
 && R_{21} = \frac{1}{u_{23}}(- c_8 v_3D_2 + c_9L_{12(3)} + c_{11}L_{23(3)})\\
 &&\mbox{} - \frac{v_2}{u_{23}}D_3L_{13(2)}^{-1}\{c_7L_{23(3)} + (c_4 - c_8)L_{12(3)} - a_{11}v_3D_2\}.
\end{eqnarray*}
The first Hamiltonian operator has the form
\begin{equation}
J_0 = \frac{1}{a_{11}u_{23}}\left(\begin{array}{cc}
    0  & 1 \\
    -1 &\displaystyle \frac{1}{a_{11}}K_{11}\frac{1}{u_{23}}
\end{array}\right)
  \label{J_0II}
\end{equation}
where $K_{12} = - a_{11}u_{23}$,
$K_{11} = a_{11}(v_2D_3 + D_2v_3) - c_4L_{13(2)} - c_7L_{23(3)} + (c_4 - c_8)L_{23(1)}$.
The corresponding Hamiltonian density reads
\begin{eqnarray}
&& H_1 = a_{11}\frac{v^2}{2}u_{23} + \frac{u}{3}\{c_9(u_{11}u_{23} - u_{12}u_{13}) + c_{11}(u_{12}u_{33} - u_{13}u_{23})\}.\nonumber\\
 \label{H1II}
\end{eqnarray}
The second Hamiltonian operator  is obtained by composing $R$ and $J_0$ as $J_1 = RJ_0$
\begin{equation}
 J_1 = \left(\begin{array}{cc}
  L_{13(2)}^{-1}             & - \left(L_{13(2)}^{-1}D_3v_2 - \displaystyle\frac{c_8}{a_{11}}\right)\displaystyle\frac{1}{u_{23}} \\
  \displaystyle\frac{1}{u_{23}}\left(v_2D_3L_{13(2)}^{-1} - \displaystyle\frac{c_8}{a_{11}}\right) & J_1^{22}
 \end{array}\right)
 \label{J_1II}
\end{equation}
with the entry $J_1^{22}$
\begin{eqnarray}
\hspace*{-15.5pt} && J_1^{22} = \frac{1}{a_{11}u_{23}}(c_9L_{12(3)} + c_{11}L_{23(3)})\frac{1}{u_{23}} - \frac{v_2}{u_{23}}D_3L_{13(2)}^{-1}D_3\frac{v_2}{u_{23}}
  \label{J_1II^22}\\
\hspace*{-15.5pt} &&\mbox{} + \frac{c_8}{a_{11}u_{23}}\left\{D_3v_2 + v_2D_3
 - \frac{1}{a_{11}}(c_4L_{13(2)} + c_7L_{23(3)} - (c_4 - c_8)L_{23(1)})\right\}\frac{1}{u_{23}}.\nonumber
\end{eqnarray}
We see that $J_1$ is manifestly skew-symmetric.

The constraint \eqref{constrH0} for the existence of the Hamiltonian density $H_0$ corresponding to $J_1$ becomes $c_{11}(c_4-c_8) = c_7c_9$. Then $H_0$ reads
\begin{eqnarray}
H_0 = -\frac{\{a_{11}(c_4 - c_8)v^2 + (a_{11}c_9u_1 + b_0)v + c_9c_8u_1^2\}u_{23}}{2\{a_{11}c_9 + c_8(c_8-c_4)\}}.
\label{H0II}
\end{eqnarray}

\subsection{Equation \eqref{eqvii}}

We show here the recursion operator and bi-Hamiltonian representation for our third example \eqref{eqvii} from section \ref{sec-integr}.
The Lax pair for this equation due to \eqref{ABvii} reads
\begin{eqnarray}
 && X_1 = \frac{\lambda}{u_{t2}}L_{23(t)} + \frac{1}{u_{t2}}(a_8L_{t1(2)} + a_{10}L_{t2(2)} + a_{11}L_{t3(2)}) \nonumber\\
 && X_2 = - \frac{\lambda}{u_{t2}}L_{t2(t)} + \frac{1}{u_{t2}}(c_7L_{t3(2)} + c_8L_{t1(2)}).
\label{Lax2}
\end{eqnarray}

In the following it is convenient to introduce the following notation
\begin{eqnarray}
&&\hat{\Delta} = a_8D_1 + a_{10}D_2 + a_{11}D_3,\quad \Delta = \hat{\Delta}[u_2]=a_8u_{12} + a_{10}u_{22} + a_{11}u_{23}\nonumber\\
&&\hat{c}=c_7D_3 + c_8D_1.
\label{not}
\end{eqnarray}

In the two component form the equation \eqref{eqvii} becomes
\begin{equation}
u_t = v,\quad v_t = \frac{q}{\Delta},\quad q = v_2(\hat{\Delta}[v] - \hat{c}[u_3]) + v_3\hat{c}[u_2].
\label{q2}
\end{equation}
From now on, square brackets denote the value of an operator.

Formulas \eqref{ABvii} also imply the recursion relations for symmetry characteristics.
\begin{eqnarray}
 && L_{23(t)}\tilde{\varphi} = (a_8L_{t1(2)} + a_{10}L_{t2(2)} + a_{11}L_{t3(2)})\varphi \nonumber\\
 && - L_{t2(t)}\tilde{\varphi} = (c_7L_{t3(2)} + c_8L_{t1(2)})\varphi.
\label{recurs2}
\end{eqnarray}

In a two-component form $u_t=v$, $\varphi_t=\psi$, $\tilde{\varphi}_t = \tilde{\psi}$ equations \eqref{recurs2} become
\[\left(\begin{array}{c}
   \tilde{\varphi} \\
  \tilde{\psi}
\end{array}\right) = R\left(
\begin{array}{c}
\varphi \\
\psi
\end{array}\right)
\]
where the recursion operator is defined by
\begin{eqnarray}
R = \left(\begin{array}{cc}
           -L_{23(t)}^{-1}v_2\hat{\Delta} & L_{23(t)}^{-1}\Delta \\
   \displaystyle -\frac{q}{v_2\Delta}D_2L_{23(t)}^{-1}v_2\hat{\Delta}+\hat{c} &\displaystyle \frac{1}{v_2}\left\{\frac{q}{\Delta}D_2L_{23(t)}^{-1}\Delta - \hat{c}[u_2]\right\}
\end{array}\right)
\label{R2}
\end{eqnarray}

The first Hamiltonian operator has the form
\begin{equation}
 J_0 =\left(\begin{array}{cc}
  0  & \Delta^{-1}\\
  - \Delta^{-1} & \Delta^{-1}K_{11}\Delta^{-1}
 \end{array}
 \right)
\label{J_02}
\end{equation}
where $K_{11}=v_2\hat{\Delta}+D_2\hat{\Delta}[v]-c_7L_{23(3)}-c_8L_{23(1)}$. With the corresponding Hamiltonian density
\begin{equation}
H_1 = \frac{v^2}{2}\Delta
\label{H1_2}
\end{equation}
the system \eqref{q2} takes the Hamiltonian form
\begin{equation}
\left(\begin{array}{c}
 u_t \\ v_t
\end{array}\right) = J_0
\left(\begin{array}{c}
 \delta_uH_1 \\ \delta_vH_1
\end{array}\right).
\label{Ham1_2}
\end{equation}

Composing the recursion operator \eqref{R2} with the first Hamiltonian operator \eqref{J_02} we obtain the second Hamiltonian operator
\begin{equation}
 J_1 = RJ_0 =\left(\begin{array}{cc}
  -L_{23(t)}^{-1}  &\displaystyle (L_{23(t)}^{-1}D_2q-\hat{c}[u_2])\frac{1}{v_2\Delta} \\[2pt]
  - \displaystyle\frac{1}{v_2\Delta}(qD_2L_{23(t)}^{-1}-\hat{c}[u_2]) & J_1^{22}
 \end{array}
 \right)
\label{J_12}
\end{equation}
where
\begin{eqnarray}
&& J_1^{22} = \hat{c}\frac{1}{\Delta} - \hat{c}[u_2]\frac{1}{\Delta}\hat{\Delta}\frac{1}{\Delta} + \frac{q}{v_2\Delta}D_2L_{23(t)}^{-1}D_2\frac{q}{v_2\Delta}
- \frac{q}{v_2\Delta}D_2\frac{\hat{c}[u_2]}{v_2\Delta}\nonumber\\
&&\mbox{} -  \frac{\hat{c}[u_2]}{v_2\Delta}D_2\frac{q}{v_2\Delta} + \frac{\hat{c}[u_2]}{v_2\Delta}L_{23(t)}\frac{\hat{c}[u_2]}{v_2\Delta}
\label{J1_22}
\end{eqnarray}
which shows that $J_1$ is manifestly skew-symmetric on account of $\Delta = \hat{\Delta}[u_2]$.

We skip here the derivation of the Hamiltonian density $H_0$ corresponding to the second Hamiltonian operator $J_1$, which is performed in a similar way to the one in section \ref{sec-biHam}, but just present the result
\begin{equation}
H_0 = kv\Delta = kv(a_8u_{12} + a_{10}u_{22} + a_{11}u_{23})
\label{H_0_2}
\end{equation}
with a constant $k$.
Thus, we obtain a bi-Hamiltonian representation for the system \eqref{q2}, which is a two-component form of the equation \eqref{eqvii}
\begin{equation}
\left(\begin{array}{c}
 u_t \\ v_t
\end{array}\right) = J_0
\left(\begin{array}{c}
 \delta_uH_1 \\ \delta_vH_1
\end{array}\right) = J_1
\left(\begin{array}{c}
 \delta_uH_0 \\ \delta_vH_0
\end{array}\right).
\label{biHam1_2}
\end{equation}

\subsection{Equation \eqref{eqviii}}

Lax pair for this equation due to \eqref{ABviii} reads
\begin{eqnarray}
 && X_1 = \frac{\lambda}{u_{t3}}L_{t3(3)} - \frac{1}{u_{t3}}L_{23(t)}\nonumber\\
 && X_2 = \frac{\lambda}{u_{t3}}(c_5L_{t2(3)} + c_8L_{t1(3)}) + \frac{1}{u_{t3}}(a_{12}L_{t3(t)} + c_6L_{13(t)} + c_7L_{23(t)}).\nonumber\\
\label{Laxiii}
\end{eqnarray}

In a two-component form, equation \eqref{eqviii} becomes
\begin{eqnarray}
&& u_t = v,\quad v_t = \frac{q}{\Delta}\nonumber\\
&& q = a_{12}v_3^2(c_5v_2u_{23} - v_3u_{22}) - c_6(v_1u_{33} - v_3u_{13})\nonumber\\
&&\mbox{} - c_7(v_2u_{33} - v_3u_{23}) - c_8(v_2u_{13} - v_3u_{12}),\quad \Delta = a_{12}u_{33}.
\label{q4}
\end{eqnarray}

First Hamiltonian operator have the form
\begin{equation}
 J_0 =\left(\begin{array}{cc}
  0  & \Delta^{-1}\\
  - \Delta^{-1} & \Delta^{-1}K_{11}\Delta^{-1}
 \end{array}
 \right)
\label{J_0iii}
\end{equation}
where $K_{11}=a_{12}(v_3D_3 + D_3v_3) -c_5L_{23(2)} - c_6L_{13(3)} - c_7L_{23(3)} - c_8L_{23(1)}$
and $K_{12} = - a_{12}u_{33} = - \Delta$. With the corresponding Hamiltonian density
\begin{equation}
H_1 = \frac{a_{12}}{2}v^2u_{33}
\label{H1_iii}
\end{equation}
the system \eqref{q4} takes the Hamiltonian form
\begin{equation}
\left(\begin{array}{c}
 u_t \\ v_t
\end{array}\right) = J_0
\left(\begin{array}{c}
 \delta_uH_1 \\ \delta_vH_1
\end{array}\right).
\label{Ham1_iii}
\end{equation}

Using the operators $A_i, B_i$ from \eqref{ABviii} we obtain the recursion relations
\begin{eqnarray*}
&& L_{t3(3)}\tilde{\varphi} = - L_{23(t)}\varphi \nonumber\\
&& (c_5L_{t2(3)} + c_8L_{t1(3)})\tilde{\varphi} = (a_{12}L_{t3(t)} + c_6L_{13(t)} + c_7L_{23(t)})\varphi.
\end{eqnarray*}
In a two-component form, with $u_t = v$, $\varphi_t = \psi$, $\tilde{\varphi}_t = \tilde{\psi}$ they become
\begin{eqnarray}
&&\tilde{\psi} = \frac{1}{u_{33}}(v_3D_3\tilde{\varphi} - L_{23(t)}\varphi)\nonumber\\
&&(c_5u_{23}+c_8u_{13})\tilde{\psi} - v_3(c_5D_2 + c_8D_1)\tilde{\varphi}\nonumber\\
&& = a_{12}v_3\psi - \frac{q}{u_{33}}D_3\varphi + (c_6L_{13(t)} + c_7L_{23(t)})\varphi.
\label{recursviii}
\end{eqnarray}
Solving the system \eqref{recursviii} for $\tilde{\varphi}$ and $\tilde{\psi}$ we obtain the recursion operator in $2\times 2$ matrix form
\[ \left(\begin{array}{c}
  \tilde{\varphi}\\
  \tilde{\psi}
 \end{array}\right) = R\left(\begin{array}{c}
  \varphi\\
  \psi
 \end{array}\right)\]
where
\begin{eqnarray}
 && R = \left(\begin{array}{cc}
 R^{11} & - a_{12}L^{-1}_{[12]3(3)}u_{33} \\[1mm]
 R^{21} &\displaystyle - a_{12}\frac{v_3}{u_{33}}D_3L^{-1}_{[12]3(3)}u_{33}
 \end{array}\right)
 \label{Riii}
\end{eqnarray}
where we introduce the notation $L_{[12]3(3)} = c_8L_{13(3)} + c_5L_{23(3)}$ and
\begin{eqnarray*}
&& R^{11} = L^{-1}_{[12]3(3)}\frac{1}{v_3}\{qD_3 - c_6u_{33}L_{13(t)} - (c_5u_{23} + c_8u_{13} + c_7u_{33})L_{23(t)}\}\\
&& R^{21} = \frac{v_3}{u_{33}}D_3L^{-1}_{[12]3(3)}\frac{1}{v_3}\{qD_3 - c_6u_{33}L_{13(t)}\\
&&\mbox{} - (c_5u_{23} + c_8u_{13} + c_7u_{33})L_{23(t)}\} - \frac{1}{u_{33}}L_{23(t)}.
\end{eqnarray*}

Composing recursion operator \eqref{Riii} with the first Hamiltonian operator \eqref{J_0iii} we obtain the second Hamiltonian operator $J_1 = RJ_0$ with the result
\begin{equation}
 J_1 = \left(\begin{array}{cc}
  - L_{[12]3(3)}^{-1}  &\displaystyle - L_{[12]3(3)}^{-1}D_3\frac{v_3}{u_{33}} \\[2pt]
  \displaystyle\frac{v_3}{u_{33}}D_3L_{[12]3(3)}^{-1} &\displaystyle - \left(\frac{v_3}{u_{33}}D_3L_{[12]3(3)}^{-1}D_3\frac{v_3}{u_{33}}
  + \frac{1}{a_{12}u_{33}}L_{23(t)}\frac{1}{u_{33}}\right)
 \end{array}
 \right)
\label{J_1iii}
\end{equation}
which shows that $J_1$ is manifestly skew-symmetric.
With the corresponding Hamiltonian density
\begin{equation}
H_0 = \{k(t,z_1)v^2 - (c_8u_1 + c_5u_2 + c_7u_3)v)\}u_{33}
\label{H0iii}
\end{equation}
the system \eqref{q4} takes the bi-Hamiltonian form
\begin{equation}
\left(\begin{array}{c}
 u_t \\ v_t
\end{array}\right) = J_0
\left(\begin{array}{c}
 \delta_uH_1 \\ \delta_vH_1
\end{array}\right)
 = J_1
\left(\begin{array}{c}
 \delta_uH_0 \\ \delta_vH_0
\end{array}\right).
\label{Ham2_iii}
\end{equation}

\subsection{Equation \eqref{eqx}}

Lax pair for this equation due to \eqref{ABx} reads
\begin{eqnarray}
 && X_1 = \frac{\lambda}{u_{t1}}L_{t1(t)} + \frac{1}{u_{t1}}(c_1L_{t1(1)} + c_3L_{t2(1)} + c_4L_{t3(1)})\nonumber\\
 && X_2 = - \frac{\lambda}{u_{t1}}L_{12(t)} + \frac{1}{u_{t1}}(a_7L_{t1(1)} + a_8L_{t2(1)} + a_9L_{t3(1)}).
\label{Lax_x}
\end{eqnarray}

In the following it is convenient to introduce the following notation
\begin{eqnarray}
&&\hat{\Delta} = a_7D_1 + a_8D_2 + a_9D_3,\quad \Delta = \hat{\Delta}[u_1]=a_7u_{11} + a_8u_{12} + a_9u_{13}\nonumber\\
&&\hat{c}=c_1D_1 + c_3D_2 + c_4D_3.
\label{notx}
\end{eqnarray}

In a two component form the equation \eqref{eqx} becomes
\begin{equation}
u_t = v,\quad v_t = \frac{q}{\Delta},\quad q = v_1(\hat{\Delta}[v] - \hat{c}[u_2]) + v_2\hat{c}[u_1].
\label{qx}
\end{equation}

First Hamiltonian operator have the form
\begin{equation}
 J_0 =\left(\begin{array}{cc}
  0  & \Delta^{-1}\\
  - \Delta^{-1} & \Delta^{-1}K_{11}\Delta^{-1}
 \end{array}
 \right)
\label{J_0x}
\end{equation}
where $K_{11} = v_1\hat{\Delta} + D_1\hat{\Delta}[v] - c_1L_{12(1)} - c_3L_{12(2)} - c_4L_{12(3)}$
and $K_{12} = - \Delta$. With the corresponding Hamiltonian density
\begin{equation}
H_1 = \frac{v^2}{2}\Delta
\label{H1_x}
\end{equation}
the system \eqref{qx} takes the Hamiltonian form
\begin{equation}
\left(\begin{array}{c}
 u_t \\ v_t
\end{array}\right) = J_0
\left(\begin{array}{c}
 \delta_uH_1 \\ \delta_vH_1
\end{array}\right).
\label{Ham1_x}
\end{equation}

Using the operators $A_i, B_i$ from \eqref{ABx} we obtain the recursion relations
\begin{eqnarray*}
&& L_{t1(t)}\tilde{\varphi} = (c_1L_{t1(1)} + c_3L_{t2(1)} + c_4L_{t3(1)})\varphi \nonumber\\
&& - L_{12(t)}\tilde{\varphi} = (a_7L_{t1(1)} + a_8L_{t2(1)} + a_9L_{t3(1)})\varphi.
\end{eqnarray*}
In a two-component form, using $u_t = v$, $\varphi_t = \psi$, $\tilde{\varphi}_t = \tilde{\psi}$, these relations become
\begin{eqnarray}
&& v_1\tilde{\psi} - \frac{q}{\Delta}D_1\tilde{\varphi} = - v_1\hat{c}\varphi + \hat{c}[u_1]\psi\nonumber\\
&& - L_{12(t)}\tilde{\varphi} = - v_1\hat{\Delta}\varphi + \Delta \psi.
\label{recursx}
\end{eqnarray}
Solving these relations for $\tilde{\varphi}$ and $\tilde{\psi}$ we obtain the recursion operator in $2\times 2$ matrix form
\[ \left(\begin{array}{c}
  \tilde{\varphi}\\
  \tilde{\psi}
 \end{array}\right) = R\left(\begin{array}{c}
  \varphi\\
  \psi
 \end{array}\right)\]
where
\begin{eqnarray}
 && R = \left(\begin{array}{cc}
 L_{12(t)}^{-1}v_1\hat{\Delta} & -  L_{12(t)}^{-1}\Delta \\[1mm]
\displaystyle \frac{q}{\Delta v_1}D_1L_{12(t)}^{-1}v_1\hat{\Delta} - \hat{c} &\displaystyle \frac{1}{v_1}\hat{c}[u_1] - \frac{q}{\Delta v_1}D_1L_{12(t)}^{-1}\Delta
 \end{array}\right).
 \label{Rx}
\end{eqnarray}
Composing the recursion operator \eqref{Rx} with the first Hamiltonian operator \eqref{J_0x} we obtain the second Hamiltonian operator
\begin{equation}
 J_1 = RJ_0 =\left(\begin{array}{cc}
  L_{12(t)}^{-1}  &\displaystyle - (L_{12(t)}^{-1}D_1q-\hat{c}[u_1])\frac{1}{v_1\Delta} \\[2pt]
  \displaystyle\frac{1}{v_1\Delta}(qD_1L_{12(t)}^{-1}-\hat{c}[u_1]) & J_1^{22}
 \end{array}
 \right)
\label{J_1x}
\end{equation}
where
\begin{eqnarray}
&& J_1^{22} = - \hat{c}\frac{1}{\Delta} + \hat{c}[u_1]\frac{1}{\Delta}\hat{\Delta}\frac{1}{\Delta} - \frac{q}{\Delta v_1}D_1L_{12(t)}^{-1}D_1\frac{q}{v_1\Delta}
+ \frac{q}{\Delta v_1}D_1\frac{\hat{c}[u_1]}{v_1\Delta}\nonumber\\
&&\mbox{} +  \frac{\hat{c}[u_1]}{\Delta v_1}D_1\frac{q}{v_1\Delta} - \frac{\hat{c}[u_1]}{\Delta v_1}L_{12(t)}\frac{\hat{c}[u_1]}{v_1\Delta}
\label{J1_22x}
\end{eqnarray}
which shows that $J_1$ is manifestly skew-symmetric on account that $\Delta = \hat{\Delta}[u_1]$.

We skip here the derivation of the Hamiltonian density $H_0$ corresponding to the second Hamiltonian operator $J_1$, which is performed in a similar way to the one in section \ref{sec-biHam}, but just present the result
\begin{equation}
H_0 = kv\Delta = kv(a_8u_{12} + a_{10}u_{22} + a_{11}u_{23})
\label{H_0_x}
\end{equation}
with a constant $k$.
Thus, we obtain a bi-Hamiltonian representation for the system \eqref{q2}, which is a two-component form of the equation \eqref{eqvii}
\begin{equation}
\left(\begin{array}{c}
 u_t \\ v_t
\end{array}\right) = J_0
\left(\begin{array}{c}
 \delta_uH_1 \\ \delta_vH_1
\end{array}\right) = J_1
\left(\begin{array}{c}
 \delta_uH_0 \\ \delta_vH_0
\end{array}\right).
\label{biHam1_x}
\end{equation}

\section{Conclusion}

We have shown that all equations of the evolutionary Hirota type in $(3+1)$ dimensions possessing a Lagrangian have the symplectic Monge--Amp\`ere form.
 We have converted the equation into a two-component evolutionary form and obtained Lagrangian for this two-component system. The Lagrangian is degenerate because the momenta cannot be inverted for the velocities.
 Applying to this degenerate Lagrangian the Dirac's theory of constraints, we have obtained a symplectic operator and its inverse, the latter being the Hamiltonian operator $J_0$. We have found the corresponding Hamiltonian density $H_1$, thus presenting our system in a Hamiltonian form.

We have developed a regular way for converting the symmetry condition to a skew-factorized form. Recursion relations and Lax pairs are obtained as immediate consequences of this representation. We have illustrated the method by the well-known heavenly equations describing self-dual gravity and produced five new explicitly integrable symplectic multi-parameter Monge--Amp\`ere equations in $(3+1)$ dimensions, as examples of the general procedure. For these equations we have derived recursion operators in a $2\times 2$ matrix form.
Composing the recursion operator $R$ with the Hamiltonian operator $J_0$, we have obtained second Hamiltonian operators $J_1=RJ_0$ for all five equations. We have found the Hamiltonian densities $H_0$ corresponding to each of $J_1$ and thereby obtained bi-Hamiltonian representations for these five systems.
Using permutations of indices and appropriate permutations of coefficients we can increase the number of explicitly integrable symplectic Monge--Amp\`ere equations and new bi-Hamiltonian systems.

\end{document}